\documentclass[twocolumn,trackchanges]{aastex63}
\usepackage{amsmath, nccmath}

\newcommand{\Cloudy}{\textsc{Cloudy}}
\providecommand{\e}[1]{\ensuremath{\times 10^{#1}}}
\providecommand{\HII}{H~{\footnotesize II}}	
\providecommand{\OIII}{[O~{\footnotesize III}]}	
\providecommand{\OII}{[O~{\footnotesize II}]}	
\providecommand{\SIII}{[S~{\footnotesize III}]}	
\providecommand{\NII}{[N~{\footnotesize II}]}	
\providecommand{\HA}{H$\alpha$}			
\providecommand{\HB}{H$\beta$}			
\providecommand{\R}{$R_{23}$}			
\providecommand{\cHB}{c$_{\text{H}\beta}$}	
\turnoffeditone
\turnoffedittwo

\shorttitle{Metal Abundances of Intermediate-Redshift AGN}
\shortauthors{Carr et al.}

\begin{document}

\title{Metal Abundances of Intermediate-Redshift AGN: Evidence for a Population of Lower-Metallicty Seyfert 2 Galaxies at z = 0.3 -- 0.4}

\correspondingauthor{David Carr}
\email{davidjcarr94@gmail.com}

\author[0000-0002-4876-5382]{David J. Carr}
\affiliation{Department of Astronomy, Indiana University, 727 East Third Street, Bloomington, IN 47405, USA}

\author[0000-0001-8483-603X]{John J. Salzer}
\affiliation{Department of Astronomy, Indiana University, 727 East Third Street, Bloomington, IN 47405, USA}

\author[0000-0001-6842-2371]{Caryl Gronwall}
\affiliation{Department of Astronomy \& Astrophysics, The Pennsylvania State University, University Park, PA 16802, USA}
\affiliation{Institute for Gravitation \& the Cosmos, Pennsylvania State University, University Park, PA 16802, USA}

\author{Anna L. Williams}
\affiliation{Department of Physics \& Astronomy, Macalester College, 1600 Grand Avenue, Saint Paul, MN  55105, USA}

\begin{abstract}

We derive oxygen abundances for two samples of Seyfert 2 (Sy2) active galactic nuclei (AGN) selected from the KPNO International Spectroscopic Survey (KISS). The two samples from KISS include 17 intermediate-redshift (0.29~$\leq$~z~$\leq$~0.42) Sy2s detected via their \OIII\ lines, and 35 low-redshift (z~$\leq$~0.1), H$\alpha$-detected Sy2s. The primary goal of this work is to explore whether the metallicity distribution of these two samples changes with redshift. To determine the oxygen abundances of the KISS galaxies, we use \Cloudy\ to create a large number of photoionization model grids by varying the temperature of the accretion disk, the ratio of X-ray to UV continuum light, the ionization parameter, the hydrogen density, and the metallicity of the narrow-line region clouds. We link the results of these models to the observed \OIII/\HB\ and \NII/\HA\ emission-line ratios of the KISS sample on the BPT diagram, interpolating across the model grids to derive metallicity. The two redshift samples overlap substantially in terms of derived metal abundances, but we find that some of the intermediate-redshift Sy2 galaxies possess lower abundances than their local universe counterparts. Our analysis provides evidence for modest levels of chemical evolution ($0.18\pm0.06$~dex) over $3-4$~Gyrs of look-back time. We compare our results to other AGN abundance derivation methods from the literature.

\end{abstract}


\section{Introduction} \label{sec:cloudyintro}

Active galactic nuclei (AGN) are the most luminous, persistent sources of light in the universe, often outshining all the stars in their host galaxies. The strong emission lines presented in AGN spectra can be used to estimate chemical abundances at a wide range of redshifts. Therefore, in order to study how the universe chemically evolves over time, it is \edit1{relevant} to study the evolution of AGN abundances with redshift.

However, AGN abundance determination methods have received little attention when compared to their star-forming galaxy (SFG) counterparts. There is a general consensus for SFGs that the determination of the electron temperature \edit1{($T_e$)} allows for the derivation of the metallicity Z \citep{2003ApJ...591..801K, 2008MNRAS.383..209H, 2012MNRAS.422..215Y, 2017PASP..129d3001P, 2019A&ARv..27....3M}. \edit1{To determine $T_e$, one needs to be able to detect weak auroral lines for atomic species in both low- and high-ionization zones or to measure the $T_e$ in one zone and then apply a scaling relation to estimate the temperature in the other zone.} This method is called the $T_e$-method. 

There are \edit1{at least} two problems with the $T_e$-method when it comes to determining AGN abundances. \edit1{Typically, for extragalactic sources, the \OIII$\lambda 4363$ auroral line is the most readily observable. However, this line is $\sim$50$-$100} times weaker than \OIII$\lambda 5007$ at the characteristic abundances of AGN, \edit1{so} it is not easily detected. \edit1{Thus,} the $T_e$-method is only applicable to objects with high-excitation lines or low metallicity, a problem of the method for both SFG and AGN abundance determinations (\citealt{1998AJ....116.2805V, 2002MNRAS.329..315C, 2007MNRAS.382..251D, 2007MNRAS.375..685P, 2016ApJ...825L..23S}, among others). 

Second, when the line is detected in AGN, the resulting electron temperature is often higher than expected and the abundances estimated from it are unrealistically low \citep{2015MNRAS.453.4102D, 2020MNRAS.492..468D}. The reasons for this are not well understood. Some authors suggest shocks enhance the \OIII$\lambda 4363$ line\edit1{. T}hough the signatures of strong shocks are not \edit1{often} found in Seyfert 2 \edit1{(Sy2)} spectra (e.g., \citealt{2013MNRAS.430.2605Z, 2015MNRAS.453.4102D})\edit1{, there is evidence that they may play a role (e.g., \citealt{2021MNRAS.501.1370D, 2021MNRAS.501L..54R})}. \citet{1997A&A...323...31K} built photoionization models of \edit1{Sy2} nuclei and considered inhomogeneities in the electron density to attempt to reproduce the narrow optical emission lines observed in these galaxies. They successfully reproduced the \SIII$\lambda \lambda 9069,9531$ emission lines, and lines like [Fe~{\footnotesize III}]$\lambda 4658$ and [Fe~{\footnotesize VII}]$\lambda 6087$, but failed to reproduce the \OIII$\lambda 4363/\lambda 5007$ line ratio. 

\citet{2020MNRAS.496.3209D} \edit1{addressed} the discrepancy between the $T_e$-method and other methods. They found that the relationship derived for \HII\ regions between temperatures of the low ($t_2$) and high ($t_3$) ionization gas does not apply to determining AGN abundances. They derived a new expression for the $t_2-t_3$ relation for \edit1{Sy2} nuclei and reduced the average discrepancy between the $T_e$-method and other abundance methods by 0.4~dex. However, a systematic difference is still present when compared to other methods.

To combat these issues, additional relationships between Z and stronger emission-line ratios have been suggested. These methods are called strong-line methods. For a review see \citealt{2019ARA&A..57..511K}. Currently there are only a handful of relationships that tie optical emission-line ratios to AGN abundance. The first relationship between metallicity and optical emission-line ratios was developed by \cite{1998AJ....115..909S}. Since then other relationships have been suggested (see \citealt{2017MNRAS.467.1507C, 2020MNRAS.492.5675C, 2020MNRAS.496.2191F, 2021MNRAS.507..466D}) but none agree perfectly with each other\edit1{, with discrepancies as large as 0.8~dex} \citep{2020MNRAS.492..468D}. Thus, if one wishes to determine the abundance of a sample of AGN with optical emission-line data, the way forward is not always clear.

In this work, we seek to derive oxygen abundances for two distinct samples of \edit1{Sy2} AGN, one at low redshifts (z $\leq$ 0.1) and the other at \edit1{higher redshifts (0.29 $\leq$ z $\leq$ 0.42), which we hereafter refer to as intermediate redshift}. Both samples of Sy2s are derived from the KPNO International Spectroscopic Survey (KISS) \citep{2000AJ....120...80S}. The primary goal of this project is to determine if the metallicity distribution of these two samples changes with redshift. 

To determine the oxygen abundances of the KISS sample of galaxies, we use \Cloudy\ \citep{2017RMxAA..53..385F} to create a set of 3240 photoionization models by varying the accretion disk temperature, ratio of X-ray to UV continuum light, ionization parameter, hydrogen density, and metallicity of the narrow-line region \edit1{(NLR)} clouds. We link the results of these models to the observed \OIII/\HB\ and \NII/\HA\ emission-line ratios of the KISS sample on the Baldwin, Philips, Terlevich (BPT; \citealt{1981PASP...93....5B}) diagram, interpolating across the model grids to derive metallicity.

Section \ref{sec:data} details the properties of the observational data from the KISS survey. Next, Section \ref{sec:method} illustrates how the photoionization models are created and how the final metallicities are derived for the Sy2 sample. Section \ref{sec:cloudyresults} presents the results of using these models to derive the abundances. Finally, Section \ref{sec:discussion} discusses these results, plots a mass-metallicity relationship, and compares this work to some of the other methods of determining AGN abundances from the literature.

\section{Observational Data} \label{sec:data}

Our sample comes from the KPNO International Spectroscopic Survey (KISS; \citealt{2000AJ....120...80S}). KISS is an objective-prism survey with optical data acquired using the Burrell Schmidt\footnote{The Burrell Schmidt telescope of the Warner and Swasey Observatory is operated by Case Western Reserve University.} telescope. It was the first purely digital objective-prism survey designed to find nearby emission-line galaxies (ELGs) using wide-field spectroscopy and its goal was to provide a comprehensive survey of ELGs in the nearby universe. The survey contains a statistically complete, emission-line-flux-limited sample of \edit1{SFGs} and \edit1{AGN} and has a limiting line flux of about $1.0 \times 10^{-15}$ erg s$^{-1}$ cm$^{-2}$. \edit1{The KISS selection criteria includes only ELG candidates with flux 5$\sigma$ above the surrounding noise.}

KISS's objective-prism spectra covered two wavelength ranges. The first, known as KISS red (KISSR), covered $6400-7200$~\AA\ and detects objects primarily by their \HA\ line \citep{2001AJ....121...66S, 2004AJ....127.1943G, 2005AJ....130.2571J}. The second, known as KISS blue (KISSB), covered $4800-5500$~\AA\ and detects objects primarily by their \OIII$\lambda$5007 line \citep{2002AJ....123.1292S}. We limit our analysis to the KISSR sample for this work because it represents \edit1{an almost seven times larger} sample of galaxies \edit1{than the sample provided by KISSB}. Furthermore, the sky overlap of the KISSB survey with the first KISSR catalog \citep{2001AJ....121...66S} results in most of the KISSB objects being included in KISSR.

While the original objective-prism spectra were enough to determine the presence of emission lines in the KISS galaxies, follow-up spectroscopy was carried out to determine accurate redshifts and the activity class of each galaxy in the KISS sample. \edit1{This spectroscopy campaign was conducted on a variety of observational facilities which include the 9.2m Hobby-Eberly Telescope (HET; \citealt{2004AJ....128..644G, 2009ApJ...695L..67S}) using the Marcario Low Resolution Spectrograph, the MDM 2.4m telescope \citep{2003AJ....125.2373W,2005AJ....130..496J} using the Mark III Spectrograph, and the WIYN 3.5m telescope \citep{2005AJ....130.2584S} using the multi-fiber positioner Hydra and the Bench Spectrograph. All available KISS spectra include \HA, \HB, \OIII$\lambda \lambda 4959, 5007$, and \NII$\lambda \lambda 6548,6584$.} See \cite{2018AJ....155...82H} for a more recent summary of the full spectroscopic follow-up of KISS.

Each galaxy in the KISSR sample also has stellar mass estimates determined using spectral energy distribution (SED) fitting \edit1{to photometric data} as described in \citet{2018AJ....155...82H}. \edit1{SEDs for each KISS galaxy were fit using the Code Investigating Galaxy Emission (CIGALE) software \citep{2009A&A...507.1793N}.} These measurements can be utilized to derive a stellar-mass-metallicity relationships which will be useful for further analysis later in this work.

\begin{figure}[t]
    \centering
    \includegraphics[width=\columnwidth,keepaspectratio]{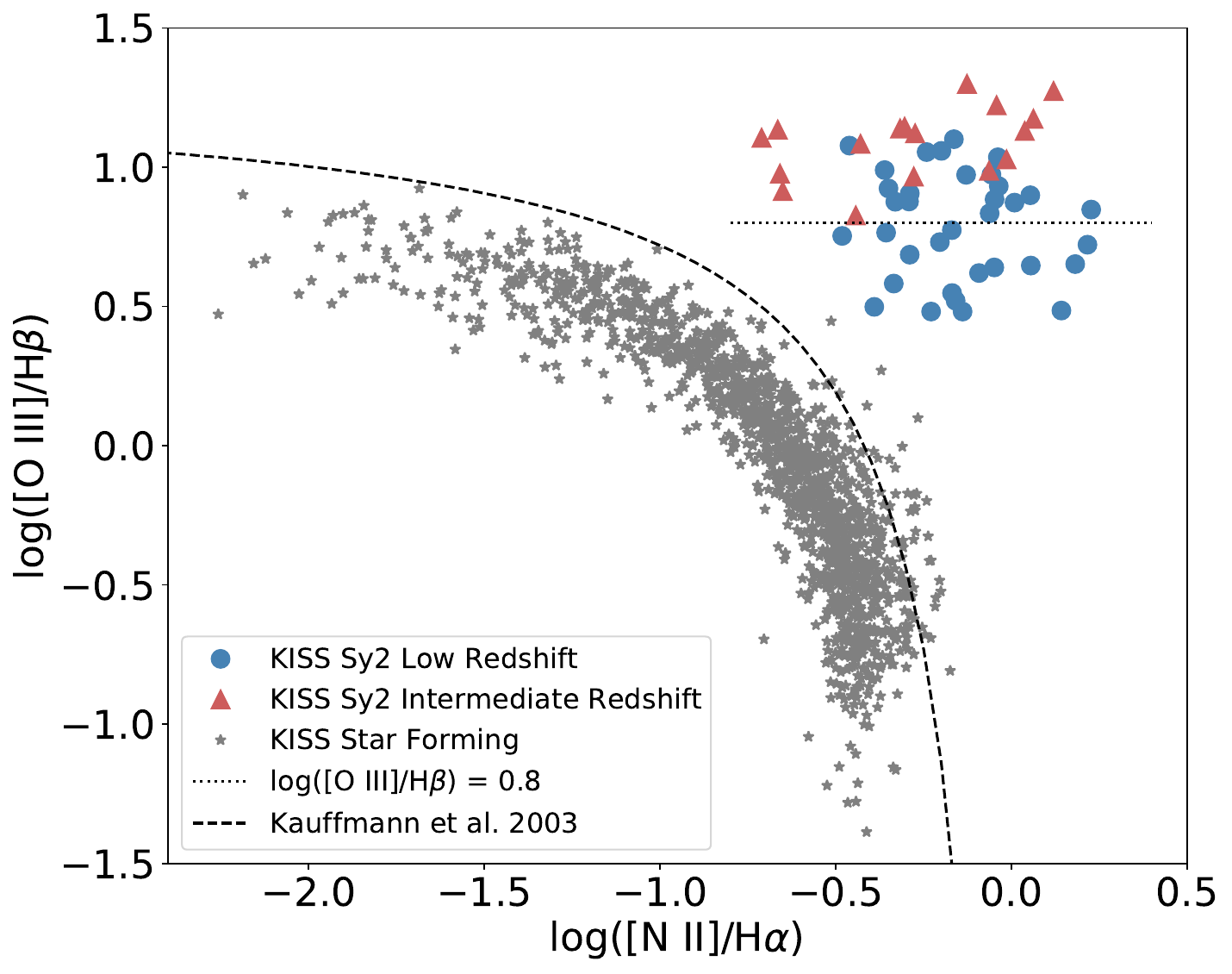}
    \includegraphics[width=0.97\columnwidth,keepaspectratio]{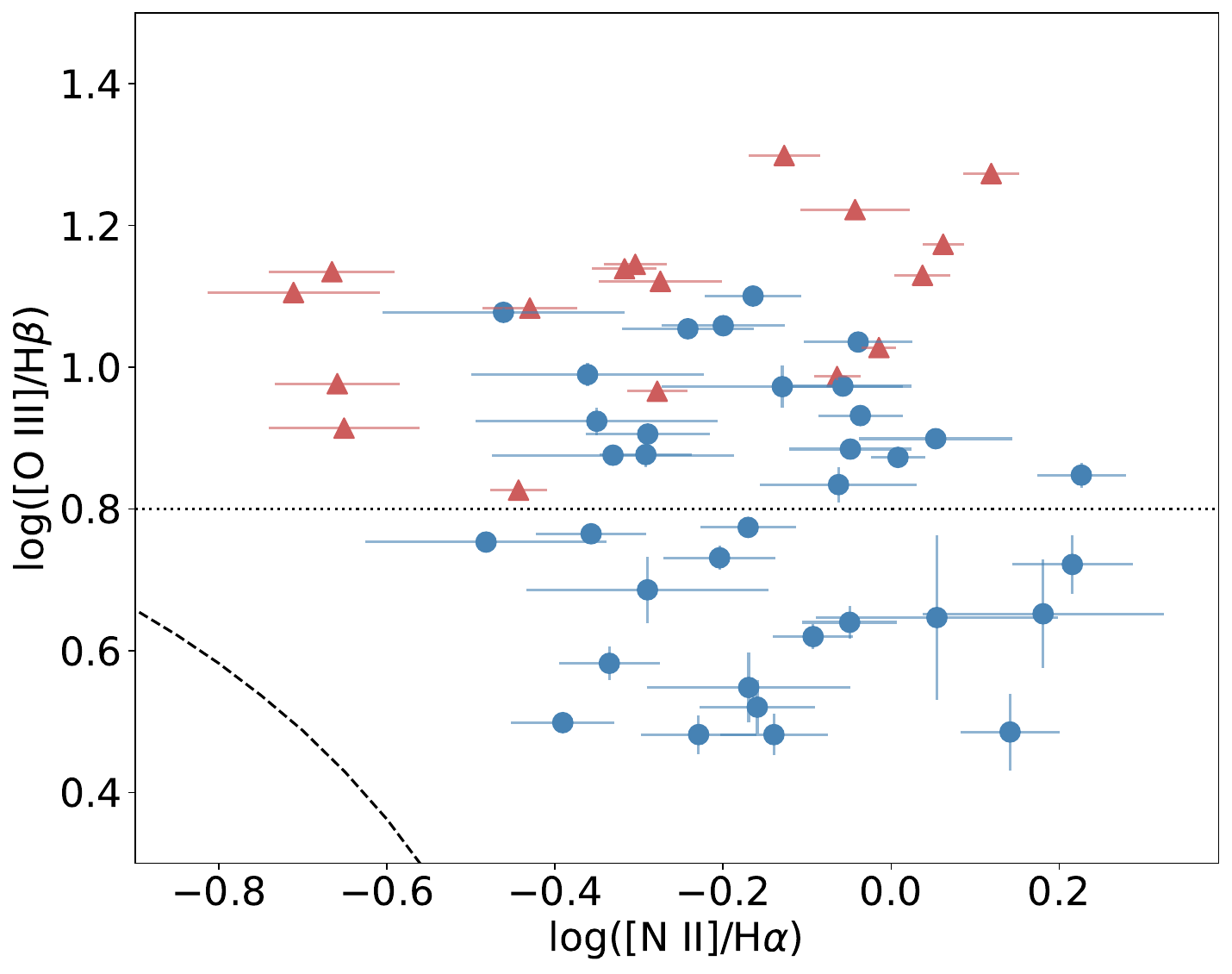}
    \caption{We present two emission-line ratio diagnostic diagrams, known as BPT diagrams \citep{1981PASP...93....5B}. 
    Top: We present the star-forming and Seyfert 2 AGN from the KISSR sample. LINERs have been excluded. The Seyfert 2 galaxies are split into two samples based on redshift and are plotted as red triangles and blue circles respectively. The star-forming galaxies plotted as gray stars. The \cite{2003MNRAS.346.1055K} line, which separates the star-forming and AGN regions of the plot, is shown as a dashed line. Finally, the horizontal dotted line at log([O~III]/H$\beta$)~=~0.8 denotes the selection limit for the intermediate-redshift Seyfert 2s, as described in the text. 
    Bottom: The Seyfert 2 region of the BPT diagram is shown in more detail. \edit1{Error bars for the emission-line ratios have been included as well as t}he \cite{2003MNRAS.346.1055K} and log([O~III]/H$\beta$)~=~0.8 lines. \label{fig:kiss_bpt}}
    \end{figure}

The KISSR star-forming and Sy2 sample are displayed on a \edit1{BPT} diagram in the top panel of Figure \ref{fig:kiss_bpt}. The empirical \cite{2003MNRAS.346.1055K} line, which separates the star-forming and AGN regions of the plot, is shown as a dashed line. Spectral classification (SFG or AGN) for the KISS sample are determined by visual inspection of \edit1{emission-line ratios and line widths}. Generally, an object is classified as a Sy2 AGN if it falls above the empirical \cite{2003MNRAS.346.1055K} demarcation line and has a log(\OIII/\HB) emission-line ratio greater than 0.5. Objects with log(\OIII/\HB) less than 0.5, while still lying above the \cite{2003MNRAS.346.1055K}, line are generally considered to be a low-ionization nuclear emission region (LINER; \citealt{1980A&A....87..152H}), as is consistent with traditional classification schemes \citep{1997ApJS..112..315H}. LINERs are excluded from the plot for clarity. 

The emission-line ratios displayed in the figure have been corrected for reddening and underlying Balmer absorption. \edit1{Internal reddening for each source was found by calculating \cHB\ (assuming \HA/\HB\ of 2.86) and the following equation \citep{2006agna.book.....O}
\begin{equation}
    \frac{I(\lambda)}{I(\text{\HB})} = \frac{I_{\lambda0}}{I_{\text{\HB}0}} 10^{-\text{\cHB} [f(\lambda) - f(\text{\HB})]}
\end{equation}
to correct the measured line ratios. Here, $f(\lambda)$ is taken from values in \cite{1982ApJ...255....1R}, which are derived from the extinction law of \cite{1958AJ.....63..201W}.} In the case of the underlying absorption, we apply a 2 \AA\ correction which is consistent with \citet{1993ApJ...411..655S} and \citet{2022ApJ...925..131H}. \edit1{This correction is purely a statistical. The mean offset for the shift in log(\OIII/\HB) is $-0.07$ and $-0.02$ for log(\NII/\HA) after this correction.}

In total, the KISSR sample contains 1455 star-forming galaxies and 52 Sy2 AGN \edit1{for a total of 1507 objects}. A majority of these objects are detected via their \HA\ emission lines at z~$\leq$~0.1. However, in $\thicksim$2\% of \edit1{these 1507} cases, objects were detected by their \OIII$\lambda$5007 emission lines that had been redshifted into the KISSR filter's wavelength range \citep{2009ApJ...695L..67S}. Many of these \OIII-detected objects are AGN and had redshifts between 0.29 and 0.42. Therefore, this work, which attempts to derive abundances for the KISS AGN, uses two distinct samples of galaxies from KISS. These samples probe different redshift windows. The first sample includes 17 \OIII-detected (0.29~$\leq$~z~$\leq$~0.42), intermediate-redshift Sy2 AGN. The second contains 35 \HA-detected (z~$\leq$~0.1), low-redshift Sy2 AGN. These are plotted in Figure \ref{fig:kiss_bpt} as red triangles and blue circles respectively. The star-forming galaxies are represented by gray stars.

It is important to note that, because the KISS sample is emission-line flux limited, it is only able to detect the strongest-lined systems via the \OIII\ line in the intermediate-redshift galaxies. This \edit1{effect biases the \OIII-detected sample and the observed limit is} $\log$\OIII$\lambda$5007/\HB~$\approx$~0.8. This limit is marked on the plot as a dotted line. We stress that this limit is strictly caused by the sensitivity limits of the survey. It is not a limit set by the physical characteristics of the galaxies themselves.

We focus in on the Sy2 sample in the bottom panel of Figure \ref{fig:kiss_bpt}. \edit1{Error bars for the emission-line ratios of each Sy2 galaxy is included in this subplot}. There appears to be an excess of intermediate-redshift galaxies with low \NII$\lambda$6584 to \HA\ ratios compared to the lower-redshift sample. \edit1{In particular, there are six intermediate-redshift Sy2s with log(\NII/\HA) less than $-0.4$ while there is only one \edit2{low-redshift galaxy} in the same region above the 0.8 limit. Further, there are 4 intermediate-redshift Sy2s with log(\NII/\HA) less than $-0.6$ while there are no low-redshift Sy2s in this region of the diagram.} \edit1{The remainder of this paper focuses on exploring the nature of these lower \NII/\HA\ objects to determine the physical cause for the lower ratio.}

\section{Methodology} \label{sec:method}

We use version 17.01 of \Cloudy\ \citep{2017RMxAA..53..385F} to create grids of AGN photoionization models that span the entire area of the KISSR sample on the BPT diagram. A metallicity is then assigned to each galaxy based on its location inside the grids. Section \ref{subsec:phot_mod} discusses the parameters specified to construct the models, Section \ref{subsec:mod_grids} presents the final model grids, and Section \ref{subsec:calc_z} outlines how a final metallicity value was assigned to each galaxy.

\subsection{Photoionization Model Parameters} \label{subsec:phot_mod}

    \begin{figure*}[t!]
    \centering
    \includegraphics[width=0.88\textwidth,keepaspectratio]{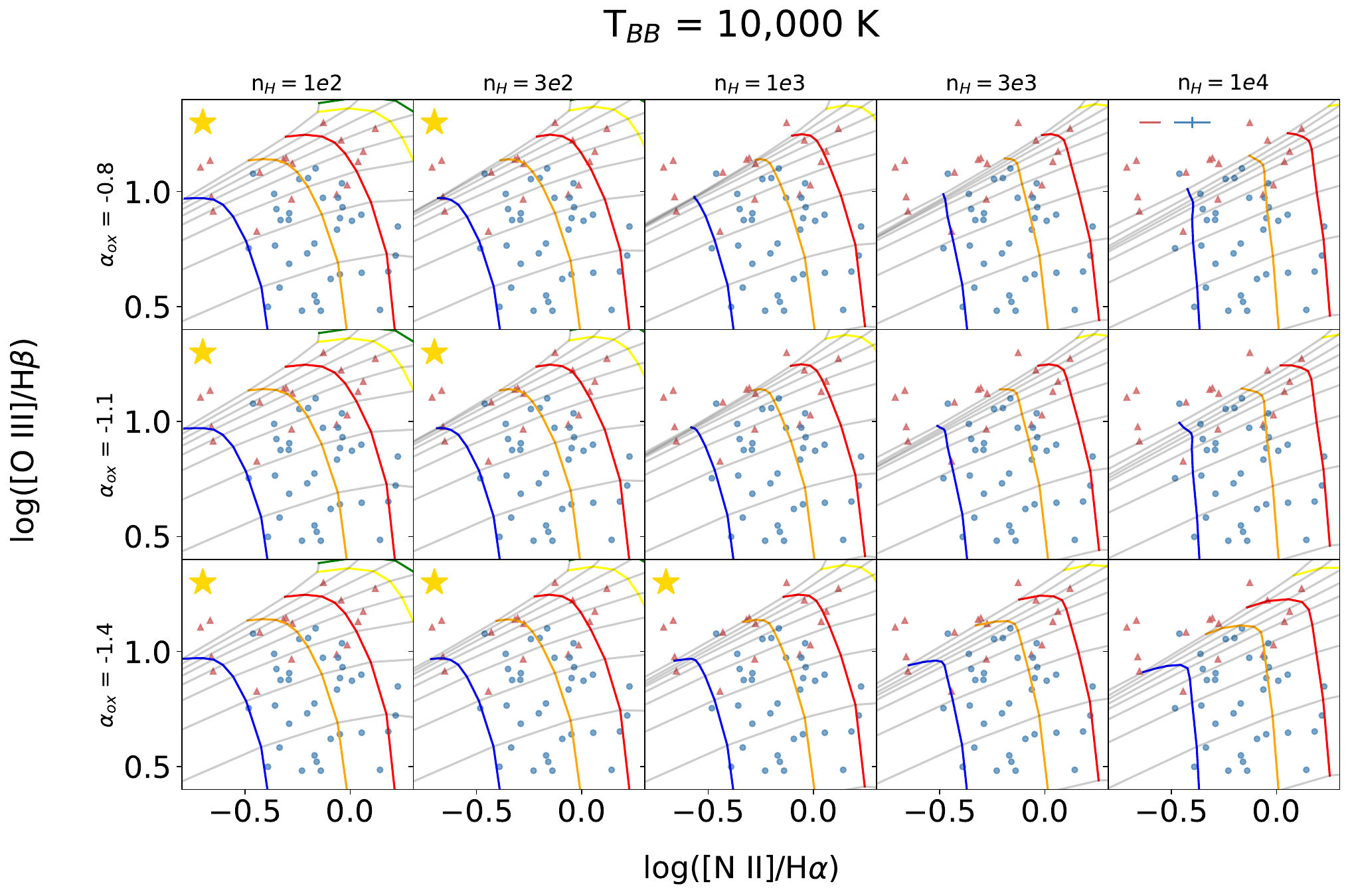} 
    \caption{We present the photoionization model grids created as part of this project overlaid on the BPT diagram of the KISSR sample. This set of 15 grids is for models calculated with $T_{BB} = 10^4$ K. Going from left to right is increasing hydrogen density and from top to bottom is decreasing $\alpha_{ox}$ values. The KISS Sy2 intermediate-redshift AGN are shown as red triangles and the low-redshift AGN are shown as blue circles as before. Each square of the group of 15 has a set of models grids represented by colored lines. Models are calculated with 2 (green), 1.5 (yellow), 1 (red), 0.75 (orange), 0.5 (blue), and 0.25 (not shown) solar metallicity. The gray lines are different ionization parameters with more negative ($-3.5$) starting to the lower right and increasing to less negative ($-1.5$) toward the upper left. \edit1{M}odels that overlay 85\% of the KISSR sample are kept to use in the final metallicity calculation. They are marked with a gold star in the upper left of the square. \edit1{Finally, average errors in the points are included in the top right subplot.} 
    \label{fig:gridbpts_1e4}
    }
    \end{figure*}
    
    \begin{figure*}[t!]
    \centering
    \includegraphics[width=0.88\textwidth,keepaspectratio]{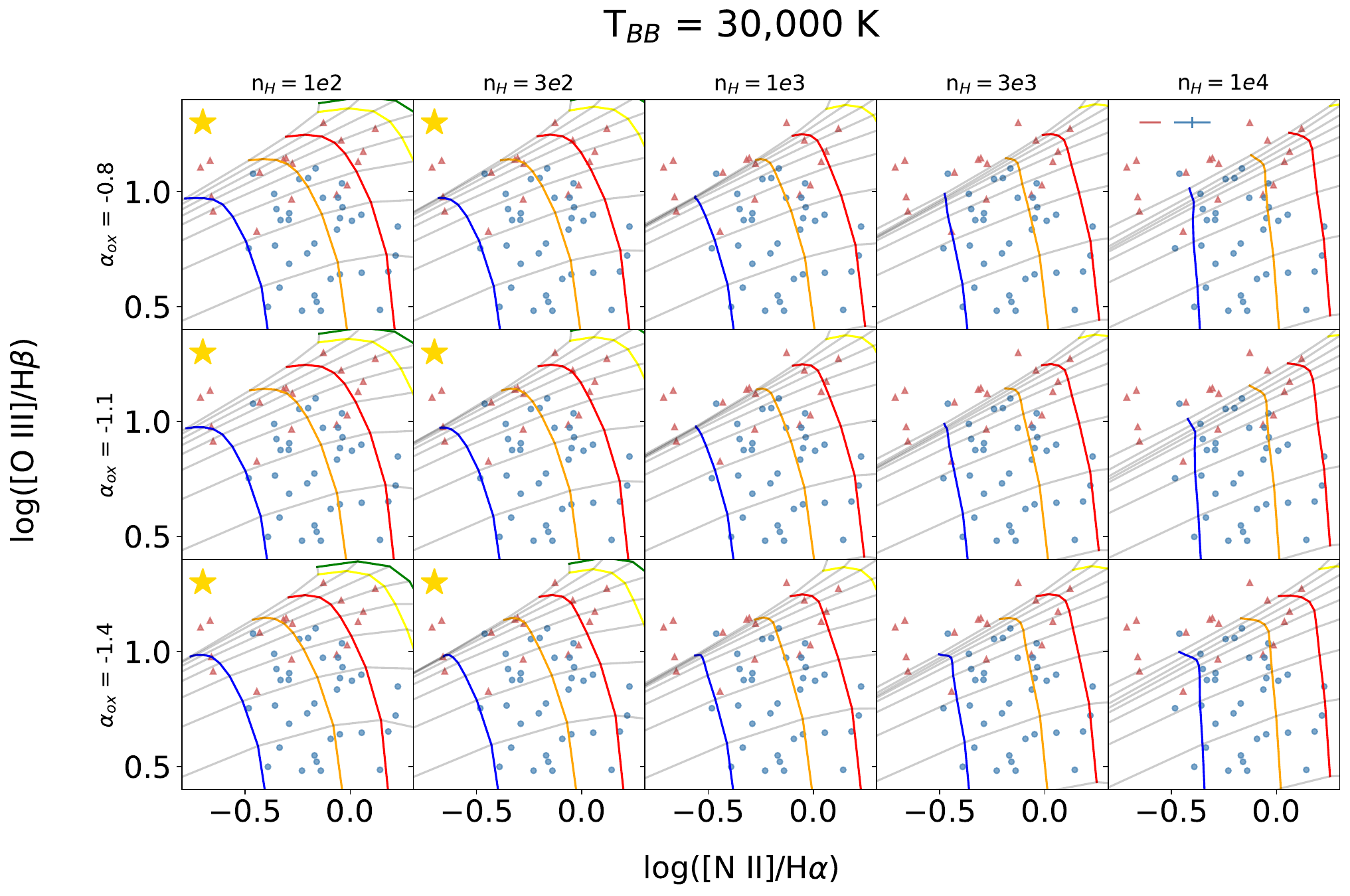}
    \caption{Similar to Figure \ref{fig:gridbpts_1e4}. This set of 15 is for grids of models calculated with $T_{BB} = 3 \times 10^4$ K.
    \label{fig:gridbpts_3e4}
    }
    \end{figure*}
    
    \begin{figure*}[b!]
    \centering
    \includegraphics[width=0.88\textwidth,keepaspectratio]{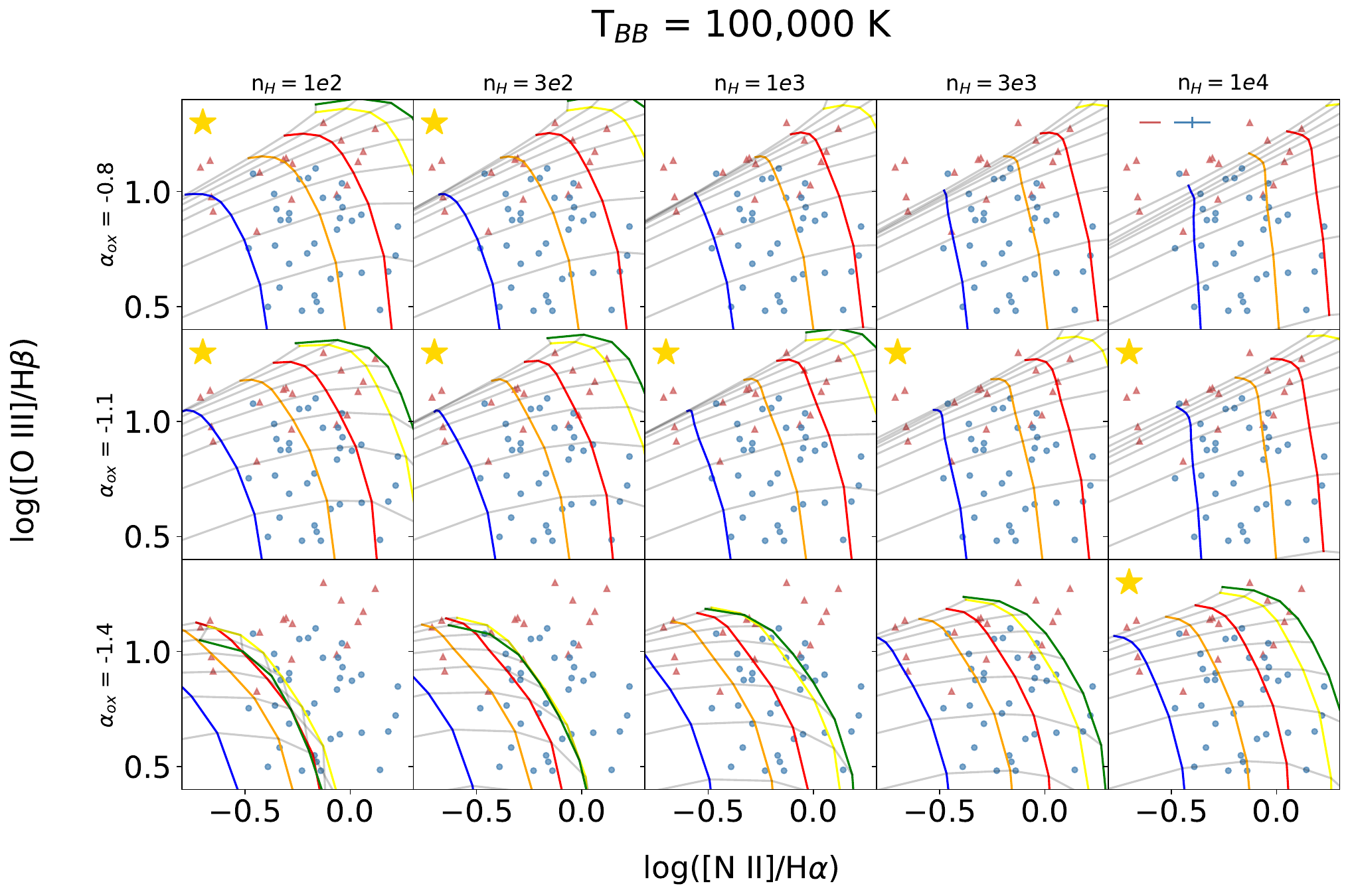}
    \caption{Similar to Figure \ref{fig:gridbpts_1e4}. This set of 15 is for grids of models calculated with $T_{BB} = 10^5$ K.
    \label{fig:gridbpts_1e5}
    }
    \end{figure*}
    
    \begin{figure*}[t!]
    \centering
    \includegraphics[width=0.88\textwidth,keepaspectratio]{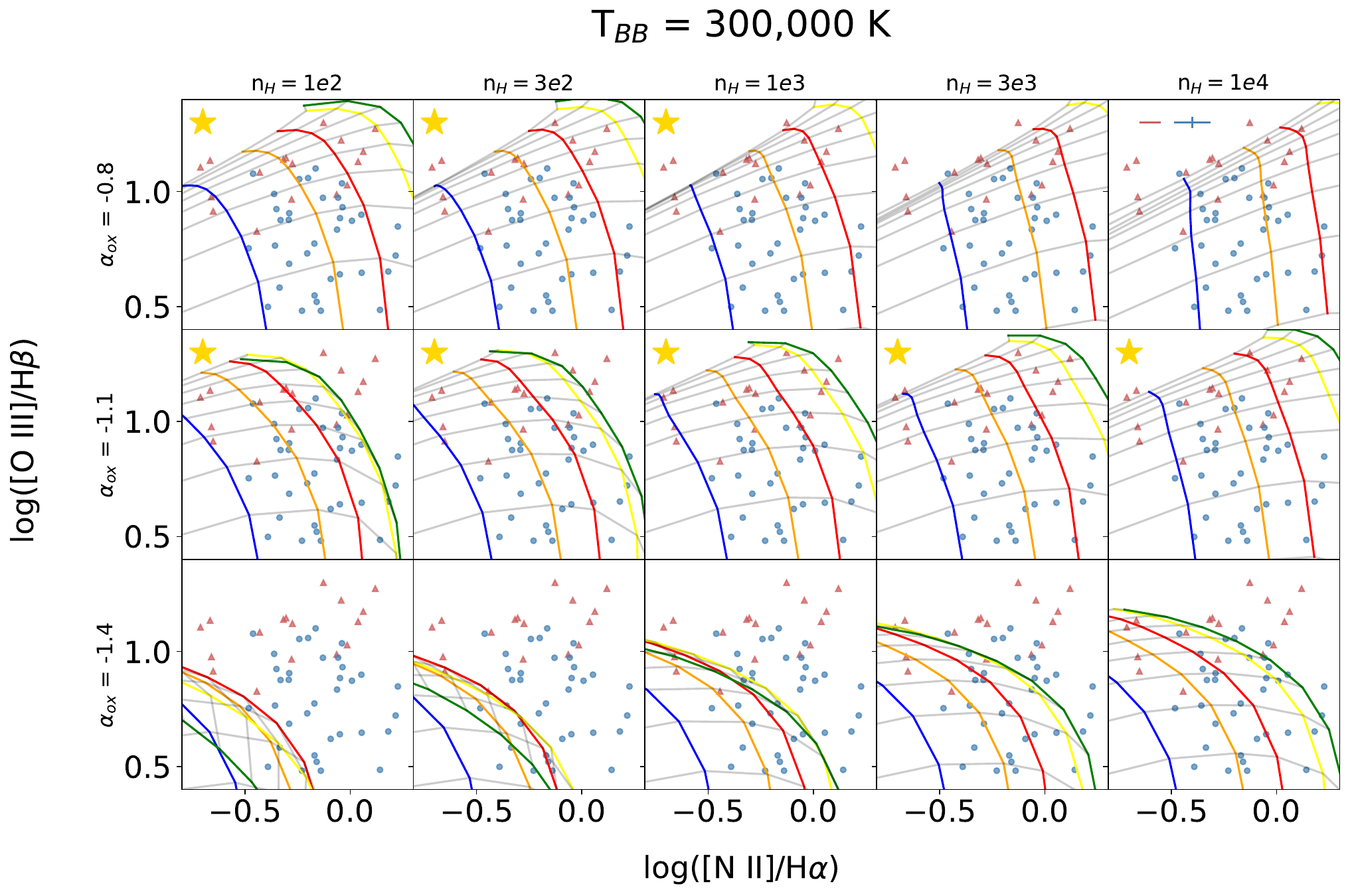}
    \caption{Similar to Figure \ref{fig:gridbpts_1e4}. This set of 15 is for grids of models calculated with $T_{BB} = 3 \times 10^5$ K.
    \label{fig:gridbpts_3e5}
    }
    \end{figure*}
    
    To create accurate approximations for the NLRs of the KISSR Sy2 sample, four main parameters had to be defined: the shape of the spectral energy distribution (SED), the ionization parameter (U), the hydrogen density (n$_H$), and the metallicity (Z).

    (i) SED: To specify the shape of the radiation field, we use \Cloudy's AGN command \edit1{which} produces a power law with two components \edit1{that has} the following form:
    \begin{equation}
        f_\nu = \nu^{\alpha_{uv}} \exp\left(\frac{-h\nu}{kT_{BB}}\right)\exp\left(\frac{-kT_{IR}}{h\nu}\right) + a\nu^{\alpha_x}
    \end{equation}
    The first component represents the ``Big Bump" (BB, sometimes referred to as ``Big Blue Bump") which peaks at 1 Ryd. The BB is thought to be a result of thermal emission from the accretion disk. \edit1{D}ue to the accretion disk's temperature varying with \edit1{distance} from the black hole, \edit1{this portion of the radiation field can be thought of as} a combination of many black bodies with different temperatures \edit1{\citep{1981ARA&A..19..137P}}.
    
    In this work we keep $\alpha_{uv}$, which represents the low-energy slope of the BB, at its default value of $-0.5$. However, we allow the value of the BB temperature ($T_{BB}$) to vary, assigning values of $10^4$, $3\times10^4$, $10^5$, and $3\times10^5$ K. \edit1{T}his range of model input values was chosen in order to cover the parameter space exhibited by the KISS emission-line ratios. Without the higher $T_{BB}$ models, two of the KISSR intermediate-redshift AGN are not covered by any of these model grids. 
    
    The second component is a power law with spectral index $\alpha_x$\edit1{. I}t represents the non-thermal X-ray radiation \edit1{and} is kept at its default value of $-1$. The X-ray to UV flux ratio, represented by $\alpha_{ox}$, is used to specify the coefficient $a$ and describes the continuum between 2 keV and 2500 \AA. In this work we vary $\alpha_{ox}$ between $-0.8$, $-1.1$, and $-1.4$. This \edit1{range of values is} used in \cite{2020MNRAS.492.5675C} and covers the range \edit1{t}ypically specified for AGN \citep{1999ApJ...516..672H, 2011ApJ...726...20M, 2019MNRAS.482.2016Z}.
    
    (ii) Ionization Parameter: The ionization parameter (U) has the functional form
    \begin{equation}
        \text{U} = \frac{\text{Q(H)}}{4\pi r^2_0 n_H c}
    \end{equation}
    and is the ratio of hydrogen-ionizing photons to total-hydrogen density. Q(H) [s$^{-1}$] is the number of ionizing photons emitted by the source \edit1{per second}, $r_0$ [cm] is the distance between the center of the source and the cloud, $n_H$ [cm$^{-3}$] is the hydrogen density, and $c$ is the speed of light.
    
    For this work, we vary the logarithm of the ionization parameter from $-3.5 \leq \log \text{U} \leq -1.5$ in 0.25~dex steps. This \edit1{range} is similar to the range used in \cite{2017MNRAS.467.1507C}, \cite{2019MNRAS.486.5853D}, and \cite{2020MNRAS.492.5675C} for AGN.
    
    (iii) The Hydrogen Density: The hydrogen density was varied between models to be $n_H = 100,\ 300,\ 1000,\ 3000,$ and $10,000$ cm$^{-3}$. Generally, the range of densities in the NLRs of AGN cover the 100 to 3000 cm$^{-3}$ range \citep{2012MNRAS.427.1266V, 2013MNRAS.430.2605Z, 2014MNRAS.443.1291D, 2020MNRAS.496.2191F} but can extend to higher densities \citep{2018ApJ...867...88R}. This range of parameters allows us to probe the entire range of the sample while staying under the critical densities for the emission lines used in this study. 
    
    \edit1{Additionally, \Cloudy\ assumes an isochoric thermal structure which we have not changed for the construction of our models.} \edit2{We carried out limited tests with} \Cloudy\ \edit2{that assumed an isobaric thermal structure. The differences between the model results are comparable to or less than the formal uncertainties in our line ratios.  Therefore, we have opted to retain the default isochoric thermal structure.}
    
    (iv) Metallicity: We allow the metallicity in the models to vary, setting it to the following values: $\text{Z}/\text{Z}_{\odot}$ = 0.25, 0.5, 0.75, 1, 1.5, and 2. \edit1{We use \Cloudy's default assumed} solar abundance of $\log(\text{O/H}) + 12 = 8.69$. Models assuming $\text{Z}/\text{Z}_{\odot} > 2$ produce similar emission-line ratios to models with metallicity less than 2 so, in order to avoid \edit1{overlap} in the model \edit1{grids}, the range was limited to $\text{Z}/\text{Z}_{\odot} \leq 2$. These values for our models are similar to previous works and have been found in AGN at a wide range of redshifts (see \cite{2020MNRAS.492.5675C} and the references therein).
    
    The abundances of all elements are scaled linearly with metallicity (O/H) except nitrogen and helium. Nitrogen was scaled with the following equation from \cite{2020MNRAS.492.5675C} based on work from \cite{2017MNRAS.468L.113D}:
    \begin{equation}
        \log(\text{N/O}) = 1.29 \times (12 + \log(\text{O/H})) - 11.84 \label{eq:N}
    \end{equation}
    \edit2{Equation \ref{eq:N}} is valid for $\text{Z}/\text{Z}_{\odot} \gtrsim 0.2$ and was derived using a sample of Seyfert 2 AGN at $z < 0.1$ and a sample of \HII\ regions. 
    
    It is necessary to introduce \edit2{a separate} scaling relation \edit2{for nitrogen} because nitrogen has two separate methods of production. For stars in low metallicity galaxies, if the oxygen and carbon are produced in the star by helium burning, then the amount of nitrogen produced is independent of the initial heavy element abundance and the nitrogen has a primary origin \edit1{\citep{2002A&A...381L..25M}}. For stars in high metallicity galaxies, the amount of nitrogen produced is proportional to the initial heavy-element abundance. It is produced from the burn of these heavy elements and is said to have a secondary origin \edit1{\citep{1993MNRAS.265..199V, 2000ApJ...541..660H}}. The relation is assumed in order to take into account the secondary nitrogen origin.
    
    For the helium abundance, a scaling relation is \edit1{also} necessary. Helium has a high initial abundance as a result of the Big Bang \edit1{and} so helium only scales weakly with metallicity. For the Helium abundance we follow a relation found in \cite{2006ApJS..167..177D}.
    \begin{equation}
        \frac{\text{He}}{\text{H}} = 0.0737 + 0.024 \times \text{Z}/\text{Z}_{\odot}
    \end{equation}

    \edit1{We do not include dust in these models following the example set by previous modeling methodologies (e,g, \citealt{2020MNRAS.492.5675C}). Dust can change the emitted spectrum of the NLR significantly by absorbing ultraviolet radiation and changing the degree of ionization of the gas, which changes the emission-line ratios observed. Furthermore, dust grain collision with gas atoms leads to more cooling while, conversely, metal depletion onto the dust grains can cause reductions in the cooling (e.g., \citealt{1992ARA&A..30...11D, 2016MNRAS.456.3354F}). \edit2{Some authors find that} models assuming the presence of dust tend not to reproduce  emission-line intensities of AGNs \citep{2006A&A...447..863N, 2009A&A...503..721M}. \edit2{Others (\citealt{2002ApJ...572..753D} and \citealt{2004ApJS..153....9G, 2004ApJS..153...75G}) suggest that the effects of dust are critical to accurately predicting AGN emission-line ratios}. \edit2{Given the lack of clear consensus on this issue, and in order to avoid} introducing \edit2{additional} uncertainty in our derived abundances, we do not include dust in these photoionization models.}

\subsection{Presentation of Model Grids} \label{subsec:mod_grids}

    The output of the models can be seen overlaid on the KISSR \edit1{Sy2} sample in Figures \ref{fig:gridbpts_1e4} through \ref{fig:gridbpts_3e5}. In total we have constructed 60 different sets of grids over a wide range of AGN parameters that cover the parameter space of the KISSR sample. Each set of grids is divided by $T_{BB}$ into groups of 15. Inside each of the groups, the hydrogen density, $n_H$, increases from left to right and the X-ray to UV flux ratio, $\alpha_{ox}$, decreases from top to bottom. The KISS Sy2 intermediate-redshift AGN are shown as red triangles and the low-redshift AGN are shown as blue circles. Each square in the group of 15 has a set of models grids represented by colored lines. These lines span the metallicity range of $\text{Z}/\text{Z}_{\odot} = 0.25 - 2$. The gray lines show the changing ionization parameter, U, with more negative ($-3.5$) starting to the lower right and increasing to less negative ($-1.5$) to the upper left. \edit1{Finally, average error bars are included in the upper right square of the group of 15 plots}. 
    
    Each of the four $T_{BB}$ groups contains 15 grids, each grid contains six metallicity values and nine ionization parameter values. In total 3240 models were created as part of this work in order to cover the full range of relevant parameter space.

    \edit1{Examining the figures, we can see that, as Z increases, \OIII/\HB\ and \NII/\HA\ also increase while, as U increases, \OIII/\HB\ increases while \NII/\HA\ decreases. Increasing $n_H$ and $\alpha_{ox}$ values pushes the model grids to slightly higher \OIII/\HB\ and \NII/\HA\ values as well. These effects are most easily seen in Figures \ref{fig:gridbpts_1e5} and \ref{fig:gridbpts_3e5}. As $T_{BB}$ increases, \OIII/\HB\ and \NII/\HA\ decrease, especially at lower $\alpha_{ox}$ values.}
    
    \edit1{Generally, models with $n_H = 100$ or $300$ cm$^{-3}$, $\alpha_{ox} = -1.1$, and $T_{BB} = 100,000$ or $300,000$ K do the best job encompassing our data, However, the only model grid that covers all but one of our data points is the $T_{BB} = 300,000$ K, $n_H = 1,000$ cm$^{-3}$, and $\alpha_{ox} = -1.1$ model grid. The $T_{BB} = 300,000$ K, $n_H = 3,000$ and $10,000$ cm$^{-3}$, and $\alpha_{ox} = -1.1$ grids in Figure \ref{fig:gridbpts_3e5} both appear to cover all of our data, but just barely miss the two, upper left intermediate-redshift AGN.}

\subsection{Model Selection and Calculating Metallicity} \label{subsec:calc_z}
    
    Since these models explore such a wide range of parameter space, some of the model grids do a better job of \edit1{aligning with} the KISSR sample than others. \edit1{The initial goal of the models was to match the observations and provide a method with which we can compare the abundances of the two samples of Sy2s. To accomplish this goal, and to ensure the models do not stray to far from the observations, we chose to limit the models used in the abundance calculation of the Sy2 samples. We decided} to only use model grids that \edit1{encompassed} 85\% of the total KISSR Sy2 sample. After we implemented this requirement, a total of 29 grids remained. These are marked with a gold star in Figures \ref{fig:gridbpts_1e4} through \ref{fig:gridbpts_3e5}.
    
   Once the \edit1{best} grids are identified, a metallicity must be calculated for each object. For each of the 29 model grids, each galaxy is located within a box-shaped region of the grid bounded by two values of the metallicity and two values of the ionization parameter. The galaxy's position is used to interpolate between the edges of the box to assign each galaxy a metallicity and ionization parameter. \edit1{This interpolation is expected to introduce an absolute grid error less than 5\% (See \citealt{2013ApJS..208...10D}).} If a point is outside the grid's area, it is not assigned a value for that set of models.
    
    After this process is done, each KISS AGN now has a number of different interpolated values for Z and U depending on where that galaxy falls in the various grids. To get the final value of Z for each AGN we take the \edit1{median and standard deviation} ($\sigma$) of the measurements. We remove \edit1{outliers} from the list of estimated metallicities \edit1{by cutting} anything that is more than 3$\sigma$ from the median, \edit1{then} recalculate the median and standard deviation, and iterate an additional time. \edit1{Using the remaining estimates, we calculate} the mean \edit1{($\mu$)} \edit1{and standard deviation} for each galaxy and set th\edit1{e mean} as its final value for Z. A similar process is done to calculate a theoretical U value for each KISSR AGN which we use later in Section \ref{sec:cloudyresults} to conduct further analysis on the KISSR sample.

    \edit1{Uncertainty in the measurement of Z is determined by varying a galaxy's position on the BPT diagram. Each galaxy is shifted by its errors in the measured emission-line ratios, then a metallicity is derived for its new position within the model grids. The average difference between its original metallicity and the shifted metallicities are added in quadrature with the recalculated standard deviation to give a final value for the maximum possible uncertainty in each galaxy's abundance.}

    \edit1{It should be noted here that there is a lot of overlap in the output of these models which makes individual metallicity measurements for galaxies in the KISSR sample suspect. Since it is impossible to know which model presents the ``right'' answer for an individual galaxy given the avaliable information from the KISSR sample, our method was chosen to cover as much parameter space as we thought reasonable, then limit the models used to those that best encompassed the observational data. Therefore, our inferred Z for any given AGN will be uncertain. However, when the method is applied to both samples of AGN, the relative metallicity distributions derived as part of this work allows for a reliable comparison of the two samples.}
    
    \edit1{We convert the oxygen abundance from $\text{Z}_{\odot}$ to $\log(\text{O}/\text{H})$ with}
    \begin{equation}
        \log(\text{O}/\text{H}) + 12 = \log(\text{Z}/\text{Z}_{\odot} \times 10^{-3.31}) + 12
        \label{eq:logohp12}
    \end{equation}
    where the solar oxygen abundance is log(O/H)~=~$-3.31$ \citep{2009ARA&A..47..481A}.

\section{Results} \label{sec:cloudyresults}
\edit1{
\begin{deluxetable*}{ccccccc}
    \tabletypesize{\scriptsize}
    \caption{Data and Results \label{tab:results_tab}}
    \tablehead{
    \\
    KISSR ID & Intermediate & log(\NII /H$\alpha$) & log(\OIII /H$\beta$) & log(M$_*$/M$_\odot$) & log(O/H)$ + 12$ & log(U) \\
    & or Low z & & & & & \\
    (1) & (2) & (3) & (4) & (5) & (6) & (7)
    }
    \startdata
  47 & int &  -0.043 $\pm$ 0.065 &   1.222 $\pm$ 0.006 &  11.39 $\pm$ 0.26 &     8.71 $\pm$  0.04 & -2.25 $\pm$ 0.08\\
 127 & int &  -0.659 $\pm$ 0.075 &   0.977 $\pm$ 0.002 &  10.60 $\pm$ 0.25 &     8.39 $\pm$  0.03 & -2.18 $\pm$ 0.32\\
 279 & int &   ~0.037 $\pm$ 0.034 &   1.130 $\pm$ 0.004 &  11.16 $\pm$ 0.09 &     8.72 $\pm$  0.04 & -2.63 $\pm$ 0.05\\
 603 & int &   ~0.062 $\pm$ 0.024 &   1.174 $\pm$ 0.005 &  12.10 $\pm$ 0.24 &     8.75 $\pm$  0.04 & -2.53 $\pm$ 0.05\\
 767 & int &  -0.015 $\pm$ 0.020 &   1.028 $\pm$ 0.003 &  11.40 $\pm$ 0.23 &     8.67 $\pm$  0.03 & -2.83 $\pm$ 0.03\\
 838 & int &  -0.278 $\pm$ 0.036 &   0.966 $\pm$ 0.003 &  11.68 $\pm$ 0.15 &     8.53 $\pm$  0.03 & -2.81 $\pm$ 0.03\\
1304 & int &  -0.127 $\pm$ 0.042 &   1.299 $\pm$ 0.003 &  11.44 $\pm$ 0.24 &     8.78 $\pm$  0.01 & -1.72 $\pm$ 0.12\\
1348 & int &  -0.304 $\pm$ 0.038 &   1.145 $\pm$ 0.002 &  10.46 $\pm$ 0.22 &     8.58 $\pm$  0.03 & -2.10 $\pm$ 0.20\\
1541 & int &  -0.444 $\pm$ 0.034 &   0.827 $\pm$ 0.003 &  11.21 $\pm$ 0.25 &     8.43 $\pm$  0.02 & -2.97 $\pm$ 0.03\\
1561 & int &  -0.275 $\pm$ 0.073 &   1.121 $\pm$ 0.005 &  11.69 $\pm$ 0.13 &     8.57 $\pm$  0.04 & -2.31 $\pm$ 0.12\\
1582 & int &  -0.317 $\pm$ 0.038 &   1.139 $\pm$ 0.009 &  11.08 $\pm$ 0.22 &     8.57 $\pm$  0.03 & -2.12 $\pm$ 0.17\\
1600 & int &  -0.430 $\pm$ 0.056 &   1.083 $\pm$ 0.002 &  11.22 $\pm$ 0.16 &     8.51 $\pm$  0.03 & -2.16 $\pm$ 0.23\\
1676 & int &  -0.651 $\pm$ 0.090 &   0.914 $\pm$ 0.002 &  10.77 $\pm$ 0.29 &     8.37 $\pm$  0.02 & -2.58 $\pm$ 0.13\\
1963 & int &  -0.064 $\pm$ 0.028 &   0.987 $\pm$ 0.005 &  10.92 $\pm$ 0.17 &     8.63 $\pm$  0.03 & -2.88 $\pm$ 0.03\\
2036 & int &   ~0.119 $\pm$ 0.034 &   1.273 $\pm$ 0.006 &  11.26 $\pm$ 0.19 &     8.82 $\pm$  0.03 & -2.26 $\pm$ 0.05\\
2097 & int &  -0.665 $\pm$ 0.075 &   1.135 $\pm$ 0.001 &  10.86 $\pm$ 0.32 &     8.48 $\pm$  0.04 & -2.02 $\pm$ 0.12\\
2363 & int &  -0.711 $\pm$ 0.103 &   1.105 $\pm$ 0.001 &  11.40 $\pm$ 0.58 &     8.44 $\pm$  0.05 & -2.13 $\pm$ 0.10\\
& & & & & & \\
   9 & low &  -0.290 $\pm$ 0.074 &   0.906 $\pm$ 0.008 &  11.09 $\pm$ 0.12 &     8.51 $\pm$  0.03 & -2.91 $\pm$ 0.04\\
 149 & low &  -0.242 $\pm$ 0.078 &   1.054 $\pm$ 0.007 &  10.15 $\pm$ 0.16 &     8.56 $\pm$  0.04 & -2.62 $\pm$ 0.06\\
 584 & low &  -0.292 $\pm$ 0.055 &   0.877 $\pm$ 0.017 &  10.96 $\pm$ 0.15 &     8.51 $\pm$  0.03 & -2.96 $\pm$ 0.03\\
 700 & low &  -0.130 $\pm$ 0.144 &   0.973 $\pm$ 0.030 &  10.78 $\pm$ 0.17 &     8.59 $\pm$  0.06 & -2.87 $\pm$ 0.10\\
 717 & low &  -0.039 $\pm$ 0.065 &   1.036 $\pm$ 0.006 &  10.28 $\pm$ 0.68 &     8.65 $\pm$  0.04 & -2.80 $\pm$ 0.06\\
 722 & low &  -0.057 $\pm$ 0.082 &   0.973 $\pm$ 0.010 &   9.77 $\pm$ 0.31 &     8.63 $\pm$  0.04 & -2.91 $\pm$ 0.07\\
 839 & low &  -0.200 $\pm$ 0.073 &   1.059 $\pm$ 0.006 &   9.93 $\pm$ 0.26 &     8.58 $\pm$  0.04 & -2.64 $\pm$ 0.06\\
1138 & low &  -0.350 $\pm$ 0.144 &   0.924 $\pm$ 0.020 &  10.20 $\pm$ 0.16 &     8.49 $\pm$  0.05 & -2.85 $\pm$ 0.07\\
1147 & low &  -0.331 $\pm$ 0.144 &   0.876 $\pm$ 0.011 &  11.47 $\pm$ 0.10 &     8.49 $\pm$  0.05 & -2.95 $\pm$ 0.07\\
1224 & low &  -0.461 $\pm$ 0.144 &   1.077 $\pm$ 0.006 &  10.59 $\pm$ 0.20 &     8.50 $\pm$  0.04 & -2.13 $\pm$ 0.24\\
1285 & low &  -0.164 $\pm$ 0.058 &   1.101 $\pm$ 0.006 &  10.86 $\pm$ 0.15 &     8.62 $\pm$  0.04 & -2.55 $\pm$ 0.06\\
2022 & low &  -0.361 $\pm$ 0.139 &   0.990 $\pm$ 0.017 &  10.62 $\pm$ 0.25 &     8.50 $\pm$  0.05 & -2.69 $\pm$ 0.09\\
& & & & & & \\
 282 & low &  -0.037 $\pm$ 0.050 &   0.932 $\pm$ 0.012 &  10.11 $\pm$ 0.17 &     8.63 $\pm$  0.03 & -2.98 $\pm$ 0.04\\
 354 & low &   ~0.053 $\pm$ 0.091 &   0.899 $\pm$ 0.014 &  10.82 $\pm$ 0.04 &     8.67 $\pm$  0.05 & -3.05 $\pm$ 0.09\\
 781 & low &   ~0.226 $\pm$ 0.053 &   0.848 $\pm$ 0.017 &  10.44 $\pm$ 0.15 &     8.77 $\pm$  0.04 & -3.12 $\pm$ 0.08\\
1556 & low &   ~0.008 $\pm$ 0.032 &   0.873 $\pm$ 0.011 &  11.08 $\pm$ 0.07 &     8.64 $\pm$  0.02 & -3.07 $\pm$ 0.03\\
1595 & low &  -0.049 $\pm$ 0.073 &   0.884 $\pm$ 0.012 &  10.32 $\pm$ 0.20 &     8.62 $\pm$  0.04 & -3.04 $\pm$ 0.06\\
2395 & low &  -0.063 $\pm$ 0.093 &   0.834 $\pm$ 0.025 &  10.74 $\pm$ 0.23 &     8.60 $\pm$  0.04 & -3.09 $\pm$ 0.07\\
& & & & & & \\
 268 & low &  -0.335 $\pm$ 0.060 &   0.582 $\pm$ 0.024 &   9.98 $\pm$ 0.36 &     8.44 $\pm$  0.03 & -3.28 $\pm$ 0.03\\
 497 & low &  -0.140 $\pm$ 0.064 &   0.482 $\pm$ 0.029 &  11.01 $\pm$ 0.03 &     8.52 $\pm$  0.02 & -3.38 $\pm$ 0.04\\
 502 & low &  -0.049 $\pm$ 0.056 &   0.640 $\pm$ 0.023 &  10.67 $\pm$ 0.07 &     8.57 $\pm$  0.03 & -3.29 $\pm$ 0.04\\
 557 & low &  -0.170 $\pm$ 0.057 &   0.774 $\pm$ 0.010 &  10.52 $\pm$ 0.22 &     8.54 $\pm$  0.02 & -3.13 $\pm$ 0.03\\
 815 & low &  -0.204 $\pm$ 0.066 &   0.731 $\pm$ 0.017 &  10.67 $\pm$ 0.17 &     8.52 $\pm$  0.03 & -3.17 $\pm$ 0.04\\
 997 & low &   ~0.142 $\pm$ 0.059 &   0.485 $\pm$ 0.054 &  10.76 $\pm$ 0.10 &     8.67 $\pm$  0.03 & -3.41 $\pm$ 0.06\\
1149 & low &  -0.290 $\pm$ 0.144 &   0.686 $\pm$ 0.047 &  11.41 $\pm$ 0.09 &     8.48 $\pm$  0.05 & -3.19 $\pm$ 0.07\\
1171 & low &  -0.482 $\pm$ 0.144 &   0.753 $\pm$ 0.010 &  11.19 $\pm$ 0.28 &     8.39 $\pm$  0.05 & -3.06 $\pm$ 0.06\\
1207 & low &   ~0.216 $\pm$ 0.072 &   0.722 $\pm$ 0.041 &  10.33 $\pm$ 0.23 &     8.74 $\pm$  0.05 & -3.25 $\pm$ 0.11\\
1219 & low &  -0.159 $\pm$ 0.069 &   0.520 $\pm$ 0.038 &  11.00 $\pm$ 0.14 &     8.52 $\pm$  0.03 & -3.35 $\pm$ 0.04\\
1321 & low &  -0.170 $\pm$ 0.121 &   0.548 $\pm$ 0.050 &  10.83 $\pm$ 0.03 &     8.52 $\pm$  0.04 & -3.33 $\pm$ 0.07\\
1382 & low &  -0.391 $\pm$ 0.062 &   0.498 $\pm$ 0.009 &  10.15 $\pm$ 0.07 &     8.40 $\pm$  0.03 & -3.33 $\pm$ 0.03\\
2073 & low &  -0.357 $\pm$ 0.065 &   0.765 $\pm$ 0.007 &  10.57 $\pm$ 0.17 &     8.46 $\pm$  0.03 & -3.08 $\pm$ 0.03\\
2129 & low &  -0.093 $\pm$ 0.048 &   0.620 $\pm$ 0.018 &  10.42 $\pm$ 0.20 &     8.55 $\pm$  0.02 & -3.30 $\pm$ 0.03\\
2311 & low &   ~0.054 $\pm$ 0.144 &   0.647 $\pm$ 0.116 &  11.11 $\pm$ 0.03 &     8.63 $\pm$  0.07 & -3.29 $\pm$ 0.14\\
2365 & low &   ~0.181 $\pm$ 0.144 &   0.652 $\pm$ 0.076 &  10.86 $\pm$ 0.10 &     8.70 $\pm$  0.08 & -3.29 $\pm$ 0.18\\
2369 & low &  -0.229 $\pm$ 0.069 &   0.482 $\pm$ 0.027 &  10.29 $\pm$ 0.17 &     8.48 $\pm$  0.03 & -3.37 $\pm$ 0.04\\
    \enddata
    \tablecomments{The low-redshift sample is split into three subsets as described in the text and Figure \ref{fig:colorbar}. The first 12 low-redshift galaxies are the galaxies with a derived ionization parameter greater than the minimum ionization parameter derived for the intermediate-redshift sample. The next six (plus the previous 12) fall above the log([O~III]/H$\beta$)~=~0.8 line. The remaining 17 fall below the log([O~III]/H$\beta$)~=~0.8 line.}
    \end{deluxetable*}
}
\edit1{We present the relevant observational data for the entire KISS Sy2 sample along with their derived log(O/H)~+~12 and log(U) values in Table \ref{tab:results_tab}. Column (1) lists the KISSR ID and column (2) indicates if the galaxy is from the intermediate- or low- redshift samples. Columns (3) and (4) show the \NII/\HA\ and \OIII/\HB\ emission-line ratios respectively and column (5) shows the log(M$_*$/M$_\odot$) values from \citet{2018AJ....155...82H}. Finally, column (6) and column (7) show the log(O/H)$ + 12$ and log(U) values derived as part of this work. Formal errors are listed for all quantities in columns (3) through (7).}

\begin{figure}
    \centering
    \includegraphics[width=\columnwidth,keepaspectratio]{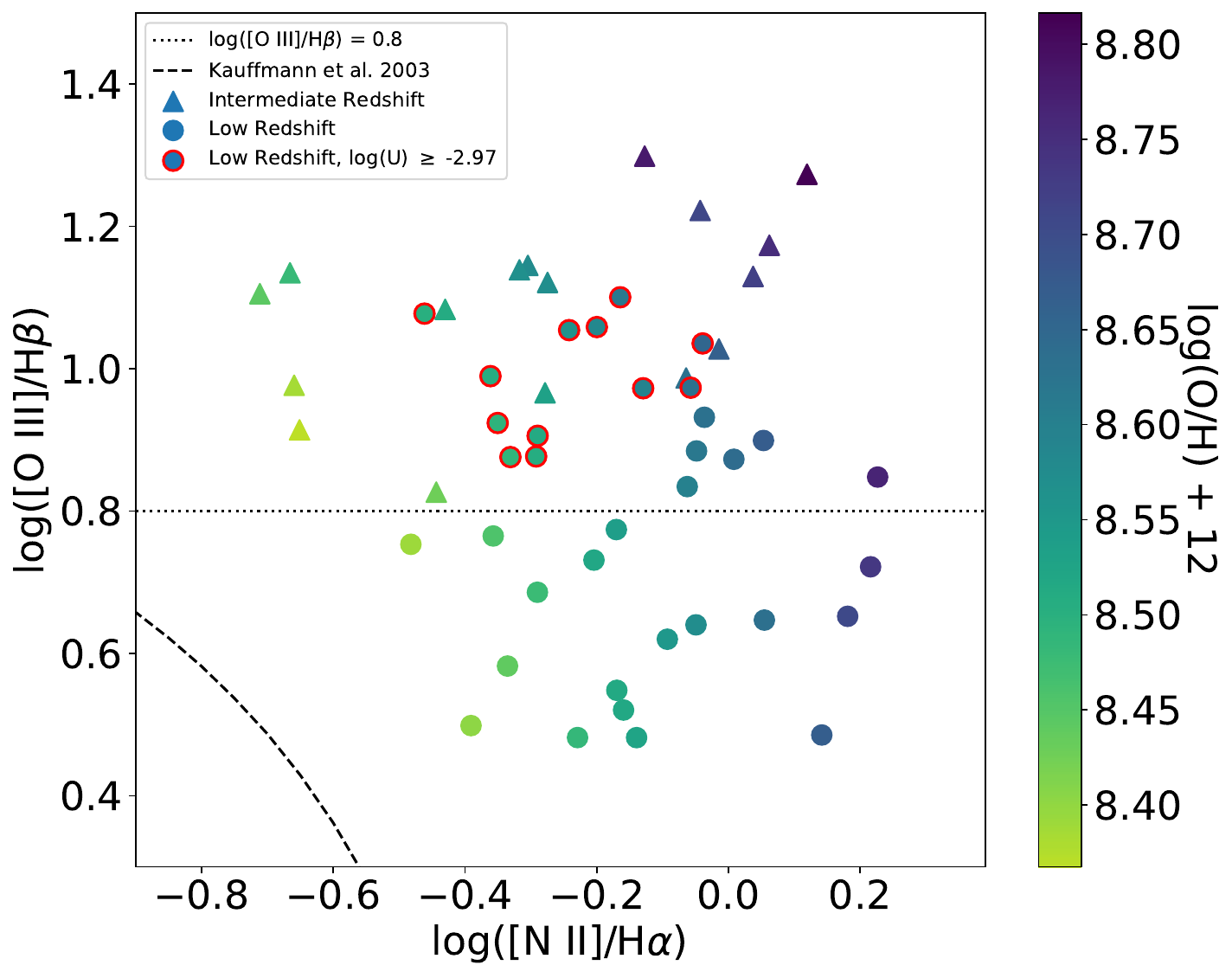}
    \caption{BPT diagram of the KISS Sy2 sample \edit1{color-coded} by their derived log(O/H)~+~12 values. The intermediate-redshift sample are triangles and the low-redshift sample are circles. The \cite{2003MNRAS.346.1055K} line is shown as a dashed line in the lower left. The horizontal dotted line at log([O~III]/H$\beta$)~=~0.8 denotes the selection limit for the intermediate-redshift Seyfert 2s, as described in the text. The low-redshift sample has been split into \edit1{several subgroups} shown in Figures \ref{fig:histO3} and \ref{fig:histU}. If the low-redshift galaxy is above the log([O~III]/H$\beta$)~=~0.8 line, it is included in Figure \ref{fig:histO3}. For a low-redshift galaxy to be included in Figure \ref{fig:histU}, it must have a derived ionization parameter, U, that is greater than the minimum ionization parameter derived for the intermediate-redshift sample. These sources are outlined in red. \label{fig:colorbar}}
    \end{figure}

\begin{figure}
    \centering
    \includegraphics[width=\columnwidth,keepaspectratio]{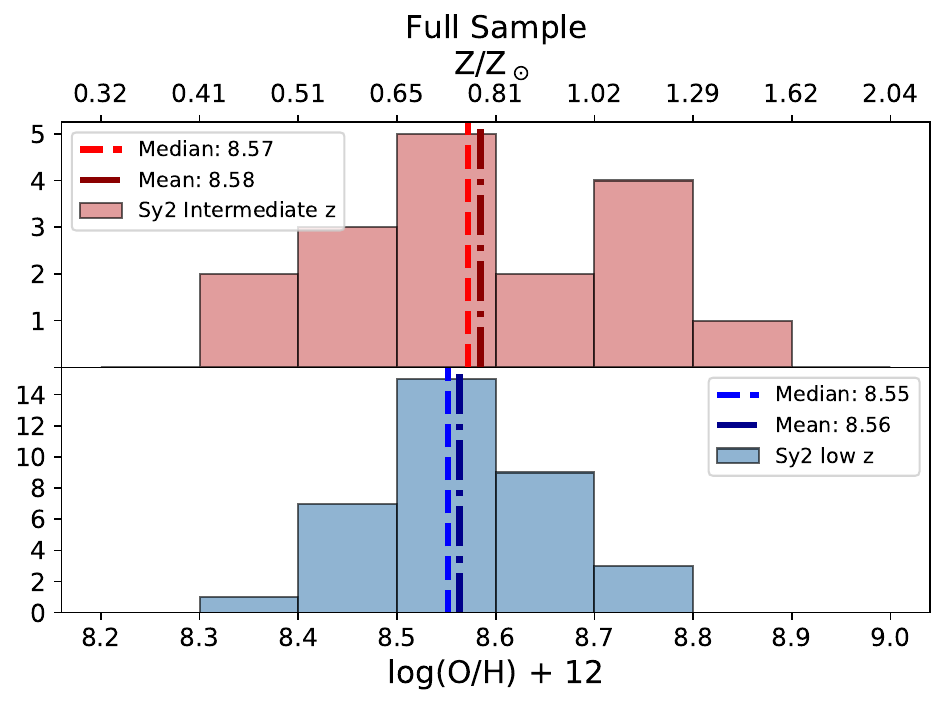}
    \caption{The full sample of KISS AGN are displayed in a histogram with the intermediate-redshift AGN in red in the top panel and low-redshift AGN in blue on the bottom panel. log(O/H)~+~12 is shown on the bottom x axis and $\text{Z}/\text{Z}_{\odot}$ is shown on the top x axis. Number is on the y axis. Median and mean values of the two samples are shown as a dashed and dashed-dotted line. \label{fig:histfullsamp}}
    \end{figure}

\edit1{W}e plot the KISS Sy2 sample on a BPT diagram in Figure \ref{fig:colorbar}. \edit1{Here} the galaxies are assigned colors based on the metallicity derived from the photoionization models. We construct a histogram of the derived abundances and plot the intermediate- and low-redshift samples separately in Figure \ref{fig:histfullsamp}. \edit1{The range of metallicity values derived by our method and presented in Figure \ref{fig:histfullsamp} are consistent with metallicity values derived through other strong-line methods (e.g., \citealt{2020MNRAS.492..468D}, Table 1). Our results span a range from log(O/H)~+~12 of 8.37 to 8.82, while the strong-line derived abundances in \citealt{2020MNRAS.492..468D} span a range of 8.39 to 9.18.}
    
\edit1{When we compare the two samples in Figure \ref{fig:histfullsamp}, t}here is essentially no difference between the mean metallicity of the intermediate- and low-redshift AGN \edit1{($\Delta \mu = 0.02$)}. However, there is a difference in the distribution of the metallicity values. \edit1{The min and max log(O/H)~+~12 values for the intermediate-redshift sample is 8.37 and 8.82 and, for the low-redshift sample, the min and max are 8.39 and 8.77. Additionally, a} large percentage of the low-redshift sample is concentrated in the 8.4 to 8.7 log(O/H)~+~12 range whereas the intermediate-redshift sample is more evenly distributed across the metallicity bins.

\edit1{I}t is not entirely \edit1{correct} to compare the entirety of these two samples together. The \HA-detected sample finds \edit1{almost} all the AGN in the volume of the survey. However, because the KISS sample is emission-line flux limited, the sensitivity of the survey limits the detection of the \OIII-selected intermediate-redshift sample to the strongest-lined systems. That is, it \edit1{preferentially} detects the AGN with high \OIII/\HB\ ratios and high ionization parameter values.

    \begin{figure}
    \centering
    \includegraphics[width=\columnwidth,keepaspectratio]{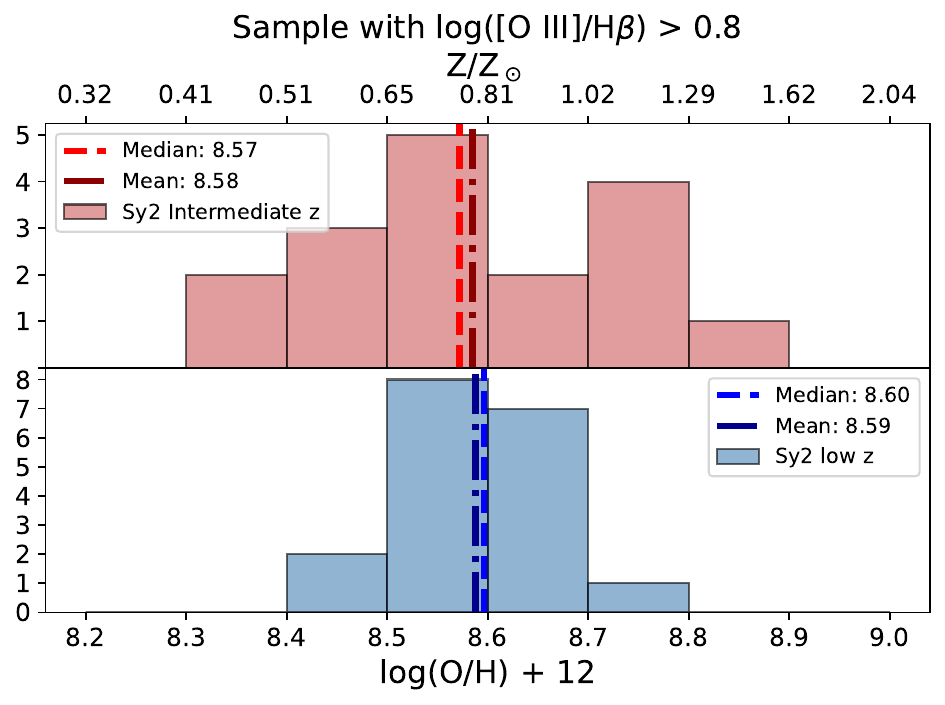}
    \caption{Similar to Figure \ref{fig:histfullsamp}, where only the low-redshift galaxies with log(\OIII/\HB)~$>$~0.8 are included. These objects are above the horizontal dotted line in Figure \ref{fig:colorbar}. The \OIII-selected, intermediate-redshift KISS Seyferts show an excess of both high- and low-metallicity objects. \label{fig:histO3}}
    \end{figure}

In order to provide a fair comparison, we can \edit1{juxtapose the intermediate-redshift sample with} two different but complementary \edit1{subsets of} the low-redshift sample. First, we \edit1{can} limit the galaxies in the low-redshift sample according to their \OIII/\HB\ ratios. The intermediate-redshift sample does not extend below log(\OIII/\HB) of about 0.8 so \edit1{we set 0.8} as the limit for the low-redshift sample. The result is shown in Figure \ref{fig:histO3}. The low-redshift points used in this comparison are above the dotted line in Figure \ref{fig:colorbar}. \edit1{Applying this limit leaves 18 of the 35 \HA-detected, low-redshift sample. The minimum log(O/H)~+~12 value in this subset is 8.49 while the max is 8.77.} 

    \begin{figure}
    \centering
    \includegraphics[width=\columnwidth,keepaspectratio]{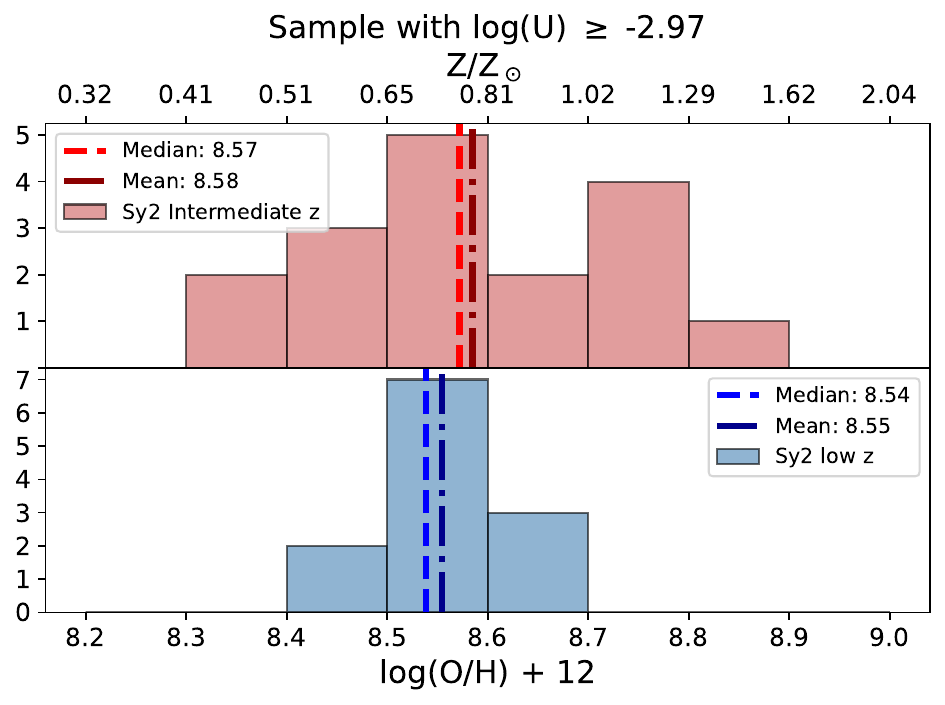}
    \caption{Similar to Figure \ref{fig:histfullsamp}, where the only low-redshift galaxies included have a derived ionization parameter, U, greater than the minimum ionization parameter calculated for the intermediate-redshift sample. These objects are outlined in red in Figure \ref{fig:colorbar}. \label{fig:histU}}
    \end{figure}

\edit1{Second,} we can \edit1{compare AGN} with similar ionization parameters. The lowest derived value of U in the intermediate-redshift sample is log(U)~$= -2.97$. The result of limiting the low-redshift sample according to the ionization parameter values is shown in Figure \ref{fig:histU} and produces similar results to Figure \ref{fig:histO3}. The low-redshift points used in this comparison are outlined in red in Figure \ref{fig:colorbar}. \edit1{Applying this limit leaves 12 of the 35 \HA-detected, low-redshift sample. The minimum log(O/H)~+~12 value in this subset is 8.49 while the max is 8.65.}

Once these effects are taken into account, we see differences in the distribution of metallicities between the two samples of galaxies, regardless of the correction method used. We highlight the following primary results. 

First, the means and medians of the two samples remain comparable and there is substantial overlap in the distribution between the two samples at intermediate abundances. However, the distribution of the intermediate-redshift AGN is much broader than that of the low-redshift AGN. \edit1{To quantify the difference in breadth, we can remove 16\% of both samples from both ends of the distributions (i.e. consider the breadth of the distributions that falls within $\pm 1\sigma$). Within this limit, the width of the intermediate-redshift sample is 0.28~dex. The corresponding widths for the two low-redshift samples are 0.14~dex for the \OIII/\HB-limited sample and 0.12~dex for the sample limited by ionization parameter.} 

Second, the intermediate-redshift sample has a higher fraction of high-metallicity AGN. \edit1{Five of the galaxies in Figure \ref{fig:histO3} in the intermediate-redshift sample have log(O/H)~+~12 value greater than 8.7 compared to one in the low-redshift subset.} The intermediate-redshift Sy2s likely include several higher mass and, therefore, higher-metallicity systems \edit1{which may contribute to this effect}. We will explore this \edit1{effect} in further detail in Section \ref{subsec:MZ}. 

Third, the intermediate-redshift sample also has a higher fraction of low metallicity AGN. \edit1{In the low-redshift subsets, no AGN has a log(O/H)~+~12 value of less than 8.49 while there are five AGN in the intermediate sample that have lower log(O/H)~+~12 values.} This implies that there is a population of Sy2s that have \edit1{somewhat} lower chemical compositions at this look-back time. \edit1{Results two and three together provide evidence for modest levels of chemical evolution over $3-4$~Gyrs of look-back time between the intermediate- and low-redshift samples.}


To test if the two distributions are different, we performed a Kolmogorov–Smirnov test \edit1{on} the two different redshift samples. For the full sample, there was a 47\% chance that the two samples were drawn from the same population \edit1{(p = 0.47)}. This percentage decreased to 37\% for the \OIII/\HB-limited sample and 27\% for the log(U)-limited sample. The results from the K-S test support the idea that there is a difference between these two samples but that the statistical significance of that difference is only modest.

    \begin{figure}
    \centering
    \includegraphics[width=\columnwidth,keepaspectratio]{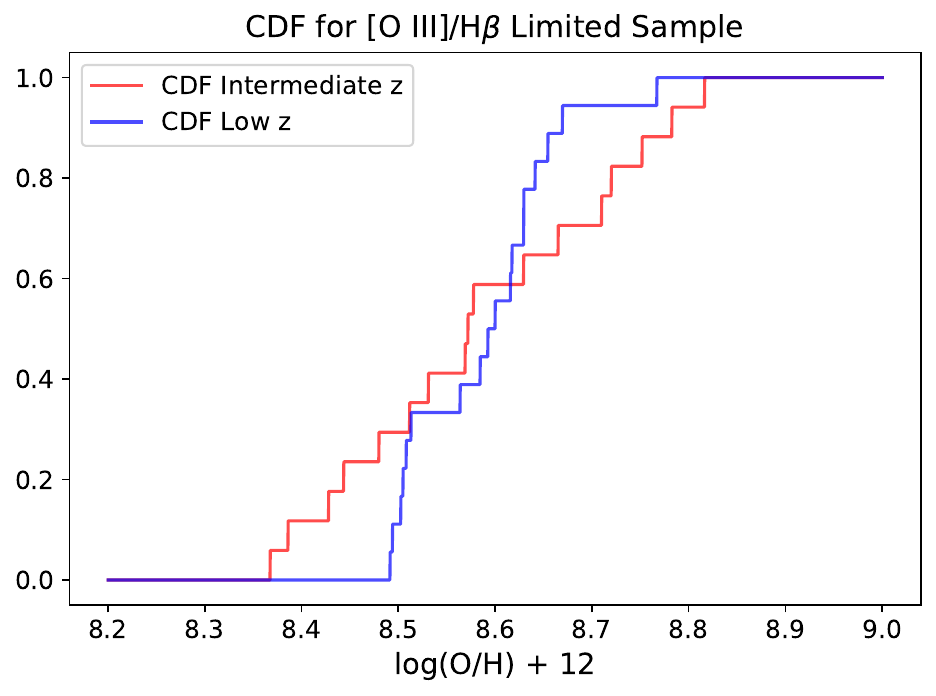}
    \caption{The cumulative distribution function associated with the intermediate-redshift sample and the subset of the low-redshift sample presented in Figure \ref{fig:histO3}. \label{fig:cdf}}
    \end{figure}

\edit1{To aid in the comparison between the two samples, we present the cumulative distribution function for the intermediate-redshift sample and the subset of the low-redshift sample described in Figure \ref{fig:histO3}. This cumulative distribution function is shown in Figure \ref{fig:cdf}.}

\section{Discussion} \label{sec:discussion}

\subsection{The Existence of a Lower-Metallicity Seyfert 2 Sample at Intermediate Redshifts} \label{subsec:lzsy2}

    Having identified a small population of \edit1{Sy2s with lower-metallicity} in our intermediate-redshift sample, we now speculate about the possible reasons for the existence of this population. It is important to note that we only see the metallicity difference in a subset of the intermediate-redshift AGN and the level of metallicity difference is \edit1{small, about} $0.1-0.2$~dex.
    
    One possible explanation for why some AGN in the intermediate-redshift sample have lower abundances is related to infalling gas from the intergalactic medium. According to cosmological models, intergalactic gas was more prevalent at higher redshifts than today (e.g., \citealt{2015ARA&A..53...51S}). If these galaxies had access to more unprocessed material, that material may flow into the centers of these galaxies. Therefore, compared to the local universe where more of the nuclear infall is from the interstellar medium of the host galaxy, the intermediate-redshift galaxies would have their metallicity content diluted. This dilution from unprocessed intergalactic gas might be what lowers the total metallicity and causes the differences between the two samples.
    
    Another reason that might explain why the intermediate-redshift sample has lower abundances is that chemical evolution in these galaxies has had less time to progress. \cite{2016ApJ...828...67L} compare the mass-metallicity relation for star-forming galaxies across three redshift bins, z~$< 0.3$, $0.3 <$~z~$< 0.5$, and $0.5 <$~z~$< 1.0$, to the mass-metallicity relation found by \cite{2013ApJ...765..140A} for z~$\sim 0.1$. They find, at a given stellar mass, an average offset of 0.13~dex at z~$< 0.3$, $-0.17$~dex at $0.3 <$~z~$< 0.5$, and $-0.24$~dex at $0.5 <$~z~$< 1.0$. Thus, the mass-metallicity relation shifts toward lower metallicity at fixed stellar mass with increasing look-back time. This \edit1{result} suggests it is possible that the higher-redshift sample has lower abundances simply because they have undergone less chemical evolution.

\subsection{Mass--Metallicity Relationship} \label{subsec:MZ}

    It is also possible that the intermediate-redshift sample includes low metallicity galaxies because these galaxies are low mass systems and lower-mass systems tend to have lower abundances (e.g., \citealt{1979A&A....80..155L}, \citealt{2018AJ....155...82H}). To test this \edit1{possibility}, we can plot the metallicity against the mass using the stellar mass data presented in \citet{2018AJ....155...82H}\footnote{The mass determination method used in \citet{2018AJ....155...82H} does not include an AGN component. Hence, the KISSR galaxy masses for both samples of Sy2 galaxies presented here are likely over-estimates. This fact should be kept in mind when interpreting our results}. Simultaneously, we can test the hypothesis that the reason for the high fraction of high-metallicity AGN in the intermediate-redshift sample is because they are higher mass and hence higher metallicity. 
    
    \begin{figure}[t]
    \centering
    \includegraphics[width=\columnwidth,keepaspectratio]{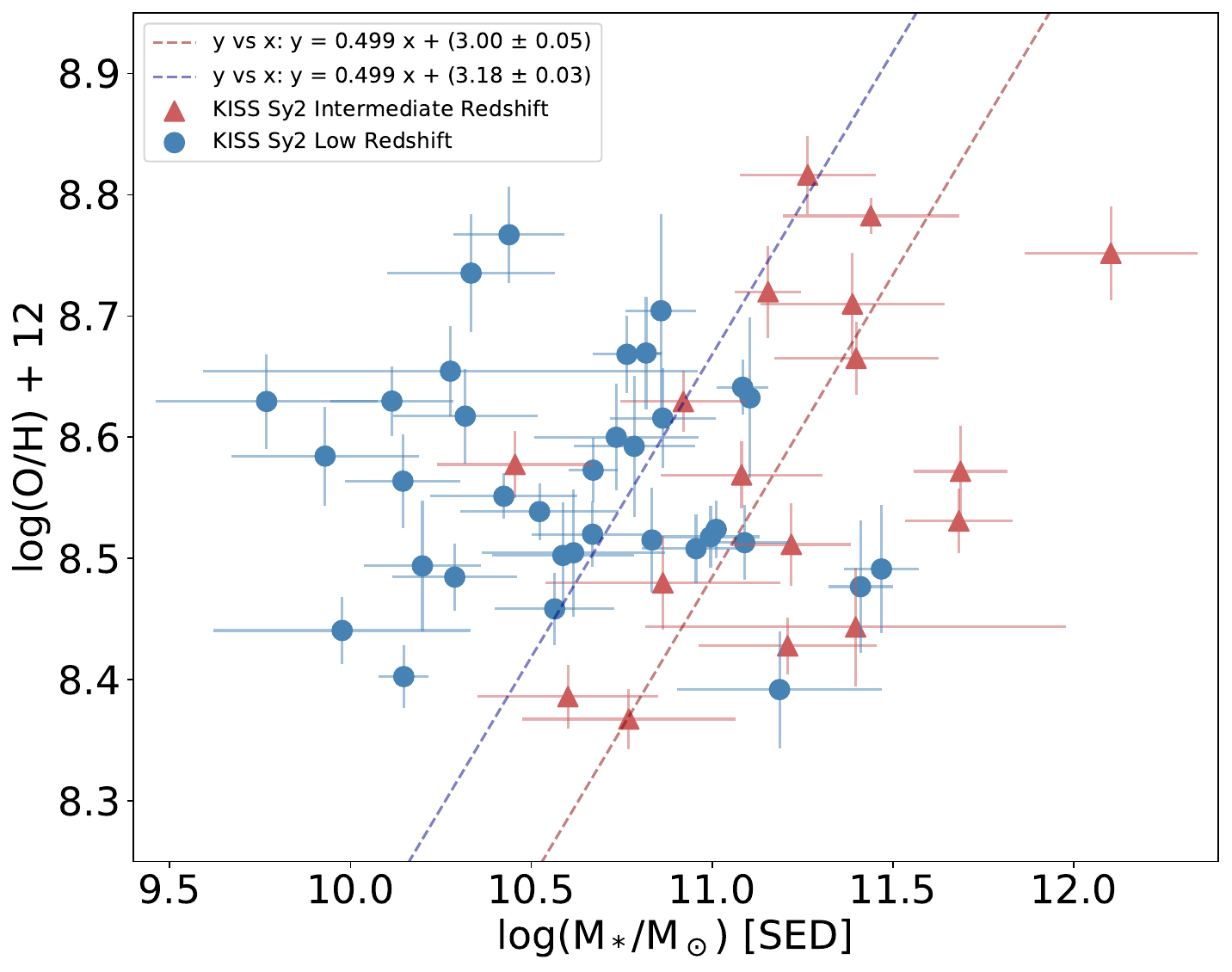}
    \caption{We plot the derived log(O/H)~+~12 values for the Sy2 samples against the stellar mass of each galaxy. The KISS Sy2 intermediate-redshift AGN are shown as red triangles and the low-redshift AGN are shown as blue circles as before. \edit1{Each sample is fit with a line with the functional form shown in the legend.}
    \label{fig:MZrelation}}
    \end{figure}
    
    Figure \ref{fig:MZrelation} shows the derived log(O/H)~+~12 values for the KISSR sample from this work plotted against the log of the stellar mass of each galaxy in a mass-metallicity (M--Z) plot. For completeness, the full low-redshift sample is displayed in the figure. The mass distribution for the low-redshift sample doesn't change significantly when we apply the \OIII/\HB\ or log(U) limits. 
    
    \edit1{There appears to be a systematic offset between the intermediate- and low-redshift Sy2 samples in Figure \ref{fig:MZrelation}. However, the limited mass range present for both samples makes fitting a formal M--Z relationship problematic. In particular, when we do this for the low-redshift sample, we get the nonphysical result that the M--Z slope is negative (i.e. higher-mass galaxies have lower abundances). As an expedient, we decide\edit2{d} to fit the two distributions using an M--Z relation with a fixed slope. For this purpose we adopt the slope of 0.499 derived by \citet{2018AJ....155...82H} for the full sample of KISS SFGs that covers a much larger mass range. We fit a M--Z relationship to the two different samples and the results are shown as straight lines in Figure \ref{fig:MZrelation}.}
    
    \edit1{The y-intercept derived from the fits changes between the two samples. The low-redshift sample's intercept is larger by $0.18\pm0.06$~dex which implies that the low-redshift sample has a higher abundance than the intermediate-redshift sample at any given mass. This metallicity offset \edit2{is in excellent agreement} with the \cite{2016ApJ...828...67L} results for SFGs in the same redshift range \edit2{(0.17 dex). The fact that galaxies with longer look-back times have lower abundances is consistent with the idea that heavier elements become more abundant as galaxies evolve over time. The observed metallicity change is only modest but appears to be clearly detected.}}
    
    \edit1{T}he \edit1{mean} value for the intermediate-redshift sample is $\log(\text{M}_* / \text{M}_{\odot}) = 11.21\pm0.10$ while the \edit1{mean} value for the low-redshift sample is $\log(\text{M}_* / \text{M}_{\odot}) = 10.63\pm0.07$. \edit1{Therefore, the intermediate-redshift sample is more massive on average by $0.58\pm0.12$~dex, a factor of $3.8$ compared to the low-redshift sample.} Of the intermediate-redshift AGN, only two fall below the \edit1{mean} of the low-redshift sample. Thus, we can say with some certainty that the low-abundance galaxies in the intermediate-redshift sample are not lower-abundance systems because of their masses. 
    
    Additionally, from Figure \ref{fig:MZrelation} we can see the higher abundance intermediate-redshift AGN are also generally higher-mass systems. This \edit1{trend} lends credence to our hypothesis that the high fraction of high-metallicity AGN in the intermediate-redshift sample is likely due to them being higher mass. The KISS survey \edit1{preferentially} detects these \edit1{Sy2} galaxies at these higher redshifts because \edit1{of their higher line luminosities which tend to correlate with higher-mass systems.}
    %

\subsection{Comparison to Previous Work} \label{subsec:comparing}

    \begin{figure*}[t]
    \centering
    \includegraphics[width=\columnwidth,keepaspectratio]{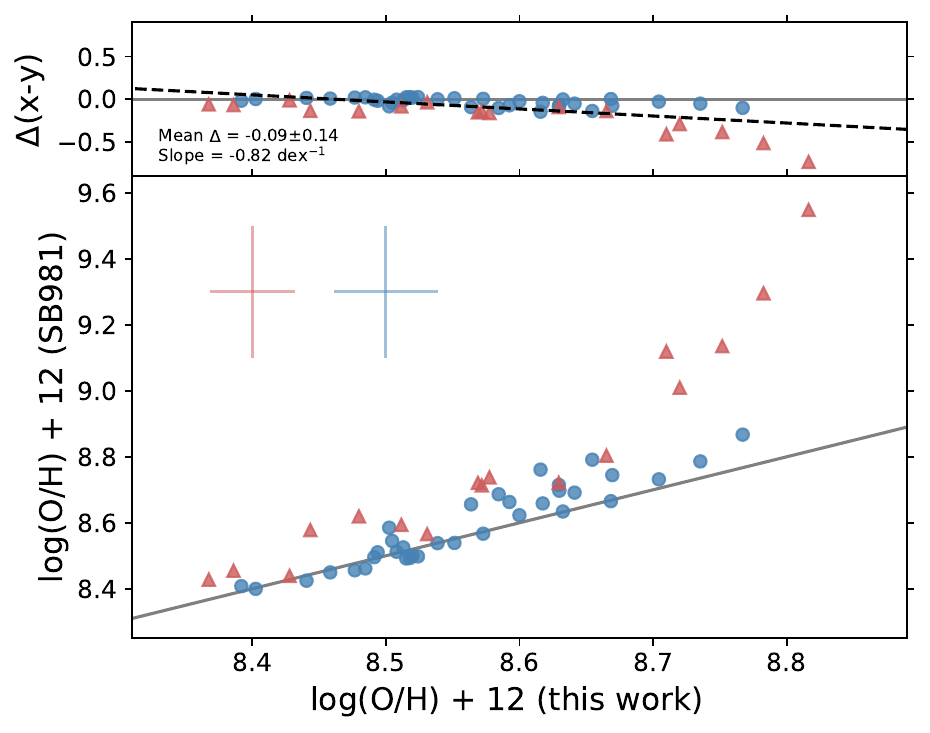}
    \includegraphics[width=\columnwidth,keepaspectratio]{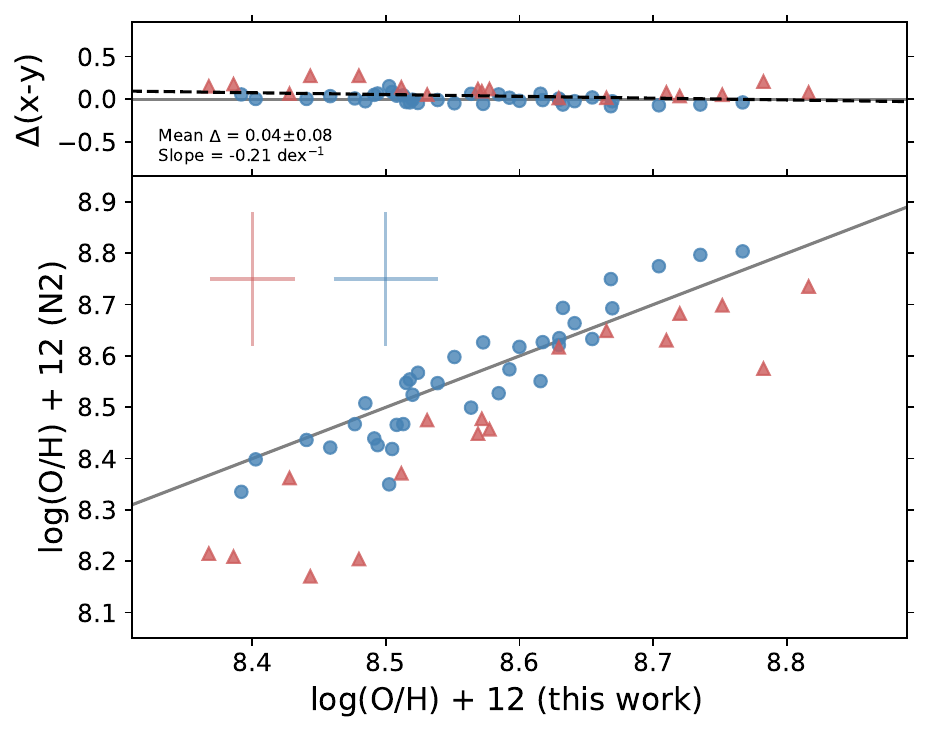}
    \includegraphics[width=\columnwidth,keepaspectratio]{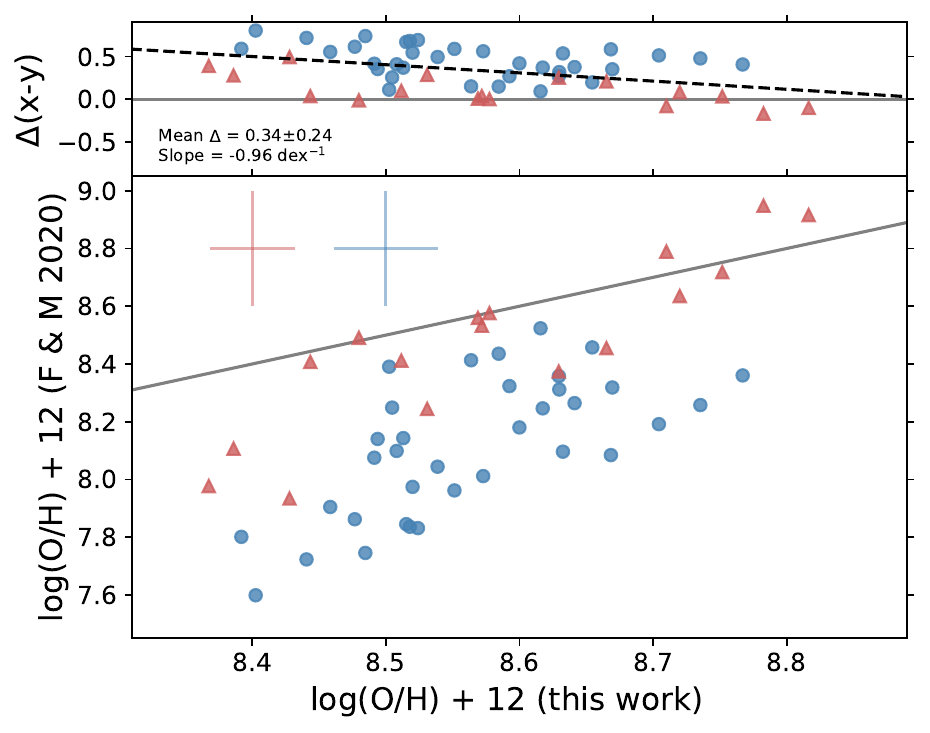}
    \caption{We plot the abundance derivations from this work against abundances derived from three different emission-line abundance methods from the literature in the bottom section of each panel. The KISS Sy2 intermediate-redshift AGN are shown as red triangles and the low-redshift AGN are shown as blue circles as before. The gray solid line is the one-to-one line \edit1{and the average errors for each sample are shown in the top left}. The top section of each panel is the difference between our results and the literature method's results ($\Delta (x-y) = \text{Our Results}-\text{Literature Results}$). The gray line in this section is a result of no difference. A line with the functional form $y = m \times x + b$ is fit to the differences and is displayed as a black dashed line. The average difference and the slope of the fit line is also listed in each panel. Left: The SB98,1 method. Right: N2 method. Lower: \cite{2020MNRAS.496.2191F} method.
    \label{fig:comparison}}
    \end{figure*}

    There are several strong-line methods from the literature that map emission-line ratios for Sy2 AGN to abundances. We can compare the results from our work to these other methods to see if they agree. Many of the relationships developed by these authors are also based on photoionization models built with \Cloudy. 
    
    The first relationship between metallicity and optical emission-line ratios was developed by \cite{1998AJ....115..909S}. They presented two different schemes which we label SB98,1 and SB98,2. The first requires \NII$\lambda \lambda 6548,6584$/\HA\ and \OIII$\lambda \lambda 4959,5007$/\HB\ line ratios and the second requires \OII$\lambda \lambda 3726,3729$/\OIII$\lambda \lambda 4959,5007$ and \NII$\lambda \lambda 6548,6584$/\HA\ line ratios. \cite{2017MNRAS.467.1507C} created the N2O2 method which is dependent on the log(\NII$\lambda 6584$/\OII$\lambda 3727$) ratio. \cite{2020MNRAS.492.5675C} created the N2 method which is only dependent on the \NII$\lambda 6584$/\HA\ line ratio. \cite{2020MNRAS.496.2191F} have developed a method for estimating abundances using the $T_e$-method. Their relationship uses the \OIII$\lambda5007$/\HB\ and \NII$\lambda 6584$/\HA\ line ratios. Finally, \cite{2021MNRAS.507..466D} created a bi-dimensional relationship using $R_{23}$ and takes advantage of the \OII$\lambda \lambda 3726,3729$, \OIII$\lambda \lambda 4959,5007$, and \HB\ lines. There are also several other approaches that make use of Bayesian statistics in the literature \citep{2014MNRAS.441.2663P, 2018ApJ...856...89T, 2019A&A...626A...9M, 2019MNRAS.489.2652P}. 
    
    In \cite{2020MNRAS.492..468D}, some of these methods were compared against each other. They found that the oxygen abundance estimates from some of the strong-line methods differed from each other by up to about $0.8$~dex with the largest differences occurring when log(O/H)~+~12~$\leq$ 8.5. However, the two \cite{1998AJ....115..909S} methods generally agree with an average difference of $-0.08$~dex.
    
    For this work, some of the KISSR sample does not include the \OII\ doublet line information. Therefore, we will focus this comparison on the methods that only use the \OIII, \NII, \HA, and \HB\ lines. Since SB98,2 relies on \OII, and because it agrees well with SB98,1, we will only consider SB98,1 going forward. We will also consider the N2 method and the \cite{2020MNRAS.496.2191F} method in this comparison. 
    
    The SB98,1 relationship is given by the following equation from \cite{1998AJ....115..909S}.
    \begin{equation}
    \begin{aligned}
    z_{\text{SB}98,1} = 8.34 + 0.212x - 0.012x^2 - 0.002y\\
    + 0.007xy - 0.002x^2y + 6.52\times10^{-4}y^2\\
    + 2.27\times10^{-4}xy^2 + 8.87\times10^{-5}x^2y^2
    \end{aligned}
    \end{equation}
    Here $z = \log(\text{O}/\text{H}) + 12$, x = \NII$\lambda \lambda 6548,6584$/\HA\, and y = \OIII$\lambda \lambda 4959,5007$/\HB. It is valid for $8.4 \leq \log(\text{O}/\text{H}) + 12 \leq 9.4$ and must be corrected for electron density effects such that
    \begin{equation}
    \begin{aligned}
        z_{\text{final}} = z_{\text{SB}98,1} - 0.1 \times \log(N_e/300 \text{ (cm}^{-1}))
    \end{aligned}
    \end{equation}
    To determine $N_e$ for each galaxy, we average the $n_H$ values from each model used to calculate the metallicity estimate for that galaxy. \edit1{Uncertainty in O/H values derived by this method were assumed to be 0.2~dex following \cite{2020MNRAS.492..468D} and \cite{2002MNRAS.330...69D}}.
    
    The N2 method provided by \cite{2020MNRAS.492.5675C} define their relationship as
    \begin{equation}
        (\text{Z}/\text{Z}_\odot) = a^{N2} + b
    \end{equation}
    where N2 = log(\NII$\lambda 6584$/\HA), a~=~4.01~$\pm$~0.08, and b~=~-0.07~$\pm$~0.01. It is valid for $0.3 \leq \text{Z}/\text{Z}_\odot \leq 2.0$ and z~$\lesssim 0.4$. \edit1{Uncertainty in the Z values derived by this method were assumed to be 30\% of the derived Z value.}
    
    Finally, \cite{2020MNRAS.496.2191F} define their relationship as
    \begin{equation}
    \begin{aligned}
        z = 7.863 + 1.170x + 0.027y -0.369x^2\\
        + 0.208y^2 - 0.406xy - 0.100x^3\\
        + 0.323y^3 + 0.354x^2y - 0.333xy^2
    \end{aligned}
    \end{equation}
    where $z = \log(\text{O}/\text{H}) + 12$, $x$ = log(\NII$\lambda 6584$/\HA), and $y$ = log(\OIII$\lambda 5007$/\HB). It is valid for $7.5 \leq \log(\text{O}/\text{H}) + 12 \leq 9.0$. \edit1{Uncertainty in the Z values derived by this method were assumed to be 0.18~dex.} 
    
    Figure \ref{fig:comparison} shows the results from the comparison of these three methods plotted against our derived abundances as well as the difference, $\Delta (x-y)$, for each method. The top left panel plots the SB98,1 method against our results. The SB98,1 method does a good job matching our results at lower abundances but, as $\log(\text{O}/\text{H}) + 12$ goes up, their results begin to deviate from ours. This \edit1{trend} is especially true for the intermediate-redshift sample, with metallicities derived by the SB98,1 relationship as high as 7.5~Z$_\odot$. Even with these large outliers, the average difference between the two methods is small, only $-0.09$. However, the slope of the line fit to the difference is comparatively large. This \edit1{exaggerated slope} is most likely due to the large differences in derived abundance in the five highest metallicity galaxies in the intermediate-redshift sample.
    
    The N2 method, shown in the top right panel, matches our derived abundances for the low-redshift sample well. However, at low and high abundances, the N2 method underestimates the intermediate-redshift sample's abundances compared to our results. Overall, this method produced results closest to ours, with an average difference of 0.04 between our results and theirs. This \edit1{result} is not a surprise, since the parameter space we explored in the models in this work is very similar to the parameter space explored by \cite{2020MNRAS.492.5675C}. 
    
    In \cite{2020MNRAS.492.5675C}, the N2 method had an average difference with the SB98,1 method of $-0.34$, whereas our results only have a $-0.09$ average difference. This \edit1{result} is a surprise considering the similar parameter spaces explored by the N2 method and our own. \cite{2020MNRAS.492.5675C} cite the reason for this difference as stemming from their (N/O)--(O/H) relationship. The (N/O)--(O/H) relationship used in their models is significantly different from the relationship used in the \cite{1998AJ....115..909S}. Our models use the same (N/O)--(O/H) relationship as the N2 method, but nevertheless produce results that match better with the results from the SB98,1 method. However, our models do explore higher hydrogen densities and $T_{BB}$ temperatures than the N2 method's photoionization models and perhaps this difference is what is causing our results to be in better agreement with the SB98,1 method. \edit1{Alternatively, this work does not examine as wide an ionization parameter space as \cite{2020MNRAS.492.5675C} and the -3.5 to -4.0 ionization parameter space explored in their work does a lot to lower their abundance measurements (see their Figure 3). This might also explain why the intermediate-redshift galaxies have worse agreement}.
    
    The comparison between our results and the \cite{2020MNRAS.496.2191F} relationship is plotted in the bottom panel of Figure \ref{fig:comparison} and show a fairly large difference in derived abundances, especially at lower metallicities. The entire low-redshift sample's abundances are underestimated by this method when compared to our results, with metallicities derived by the \cite{2020MNRAS.496.2191F} relationship that are as low as 0.08 Z$_\odot$. Such low Z values seem unphysical given the stellar masses of these systems (see Section \ref{subsec:MZ}). However, the intermediate-redshift sample is matched fairly well by this relationship at intermediate and higher abundances. 
    
    The \cite{2020MNRAS.496.2191F} method produced the largest difference when compared to our results with an average difference of 0.34. Initially, we thought the difference between our abundances was due to the two samples being located at different redshifts, but the median redshift of the parent sample from \cite{2020MNRAS.496.2191F} is 0.075 which is not so different than the median of the KISSR low-redshift sample of z~=~0.063 \citep{2003AJ....125.2373W, 2004AJ....128..644G, 2005AJ....130..496J, 2005AJ....130.2584S, 2009ApJ...695L..67S}. The deviation of the \cite{2020MNRAS.496.2191F} method from our method is similar to discrepancies seen between strong-line methods and the $T_e$-method \citep{2020MNRAS.492..468D}. These large deviations are typically at lower abundances and are the signature of a historical problem that results from using the $T_e$-method for Sy2 abundances (discussed in Section \ref{sec:cloudyintro}).

\section{Summary and Conclusions} \label{sec:cloudysummary}

It is important to study how AGN change over time to better understand how galaxies, and the universe, change and evolve with redshift. AGN abundance determination is still an area of active study. In this work, we derive the abundances for two samples of Sy2 AGN from the KISS survey. One of these samples is at low redshifts of z $\leq$ 0.1 and the other is at intermediate redshifts of 0.29 $\leq$ z $\leq$ 0.42. We create a large number of photoionization models to \edit1{encompass} the parameter space exhibited by the KISS emission-line ratios and overlay them on the BPT diagram with the KISS sample. We interpolate within the model grids to derive abundances for the KISS AGN. 

When the differences between the two samples are taken into account, we can draw three conclusions. First, we find that the \edit1{breadth of the} distribution of the intermediate-redshift AGN abundances is broader than that of the low-redshift AGN \edit1{(0.28~dex compared to about 0.14~dex)}. Second, the intermediate-redshift sample has \edit1{an approximately five times} higher fraction of high-metallicity AGN. This \edit1{higher fraction} is probably because many of the higher-redshift Seyferts are higher-mass systems and hence have higher metallicity. Third, the intermediate-redshift sample also has a higher fraction of lower-metallicity AGN. This \edit1{higher fraction} implies that there is a population of Sy2s that have \edit1{modestly} lower ($0.18\pm0.06$~dex) chemical compositions at this look-back time. These conclusions are further confirmed when the Sy2 sample is plotted on a mass-metallicity relationship diagram. We speculate that the reason for this lower-abundance population of Sy2s at higher redshifts could be due to either the inflow of unprocessed intergalactic gas or because the higher-redshift sample has had less time to chemically evolve.

The results from this work are compared to three different strong-line methods from the literature. These comparisons present mixed results. The abundances derived for the low-redshift sample are in good agreement with two of the methods from the literature: the SB98,1 method and the N2 method. For the intermediate-redshift sample, depending on which method is compared, some of the derived abundances are in good agreement while others show large differences. The SB98,1 method is generally able to match our results for the higher-redshift sample at lower and intermediate abundances, the N2 method does best at intermediate abundances, and the \cite{2020MNRAS.496.2191F} method matches our results best at intermediate and higher abundances. Regardless of these disagreements, the breadth of parameter space covered by our models should allow this methodology to be applied to other Sy2 samples in the future and enable the derivation of AGN abundances in future studies.

\begin{acknowledgments}

We would like to thank the College of Arts and Sciences at Indiana University (IU), the IU Department of Astronomy, and the IU Graduate and Professional Student Government who helped to support this project by providing funds for participation in various meetings to present this work. We'd like to thank Dr. Gary Ferland, Dr. Oli Dors, and Dr. Christopher Agostino for their useful and helpful comments regarding the photoionization models. The early phases of the KISS project were supported by NSF grants AST 95-95020 and AST 00-71114 to J.J.S.  Many colleagues from the KISS Team contributed to the data acquisition and analysis that made this project possible. Special thanks to Anna Jangren, Jessica Werk, Laura Chomiuk, Joanna Taylor, Duane Lee, Janice Lee, Jason Melbourne, Matt Johnson, Alec Hirschauer, and Steven Janowiecki. \edit1{We would also like to thank the referee for their extensive and insightful comments which have significantly improved this paper.}

\end{acknowledgments}

%



\software{\Cloudy\ \citep{2017RMxAA..53..385F}} 

\bibliography{mybib}{}
\bibliographystyle{aasjournal}

\end{document}